# Molecular Hydrogen in Star-forming regions: implementation of its micro-physics in Cloudy

G. Shaw,[1] G. J. Ferland,[1] N. P. Abel,[1] P. C. Stancil,[2] and P. A. M. van Hoof[3]

## Abstract

Much of the baryonic matter in the Universe is in the form of $H_2$ which includes most of the gas in galactic and extragalactic interstellar clouds. Molecular hydrogen plays a significant role in establishing the thermal balance in many astrophysical environments and can be important as a spectral diagnostic of the gas. Modeling and interpretation of observations of such environments requires a quantitatively complete and accurate treatment of $H_2$. Using this micro-physical model of $H_2$, illustrative calculations of prototypical astrophysical environments are presented. This work forms the foundation for future investigations of these and other environments where $H_2$ is an important constituent.

*Subject headings:* ISM : molecules --- PDR

---

[1] University of Kentucky, Department of Physics and Astronomy, Lexington, KY 40506; gshaw@pa.uky.edu, gary@pa.uky.edu, npabel2@uky.edu

[2] University of Georgia, Department of Physics and Astronomy and Center for Simulational Physics, Athens, GA 30602; stancil@physast.uga.edu

[3] APS Division, Department of Physics and Astronomy, Queen's University Belfast, BT7 1NN, Northern Ireland; p.van-hoof@qub.ac.uk

# 1 Introduction

Most of the hydrogen in the interstellar medium is in the form of $H_2$ (Field et al. 1966; Shull & Beckwith 1982; Black & van Dishoeck 1987; Draine & Bertoldi 1996). However, $H_2$ is not an efficient emitter due to the lack of a permanent electric dipole moment and its small moment of inertia, resulting in large energy level spacings. $H_2$ helps to define the chemical state of the gas, it plays a major role in exciting important gas coolants, and it comprises the majority of the mass. Therefore complete numerical simulations of a non-equilibrium gas must include a detailed treatment of the physics of $H_2$.

This paper describes the implementation of an extensive model of the $H_2$ molecule into the spectral simulation code Cloudy. The code was last described by Ferland et al. (1998). Cloudy determines the ionization and excitation state of all constituents self-consistently by balancing all ionization and excitation processes, the electron density from the ionization structure, and the gas kinetic temperature from the balance between heating and cooling processes (e.g. Osterbrock & Ferland 2005). This approach, as discussed in detail by Ferland (2003), has energy conservation as its foundation. The goal of the current implementation is to have all interactions of $H_2$ with its environment, both collisional and radiative, determined self-consistently with a limited number of free parameters. Only the gas composition, density, column density, dynamical state, and incident radiation field must be specified to uniquely determine the physical conditions and emitted spectrum.

# 2 Physical model of the $H_2$ molecule

## 2.1 Spectroscopic notation and energy levels

Figures 1, 2, and 3 show the energy levels included in our model. The figures show, respectively, all energy levels, the levels within the ground electronic state,

and the lower levels within this state. There are 1893 levels in the model which together produce a total of 524,387 lines.

Energies for the 301 rovibrational levels within the $1s\ ^1\Sigma_g$ (denoted as X) ground electronic state[4] were adopted from the experimental compilation of Dabrowski (1984). Errors in the energies of eight of the levels have been corrected and energies of 17 missing levels have also been included (Roueff 2004, private communication).

We also include the rovibrational levels within the lowest 6 electronic excited states that are coupled to the ground electronic state by permitted electronic transitions. The transition energies of Abgrall et al. (2000) were used to obtain the upper electronic state rovibrational energies. These configurations and shorthand notation are $2p\ ^1\Sigma_u^+$ (B), $2p\ ^1\Pi_u$ (C+ and C-), $3p\ ^1\Sigma_u^+$ (B′), and $3p\ ^1\Pi_u$ (D+ and D-). These electronic excited states are relevant because they participate in the Solomon process, the dominant $H_2$ destruction mechanism when radiation is present in the wavelength region between the Lyman continuum and Ly$\alpha$. Roughly 10% of the excitations following continuum radiation through the electronic lines are followed by decays into the X continuum (Abgrall et al. 1992). The majority of these photoexcitations are an indirect source for populating highly excited X rovibrational levels which decay to lower levels producing infrared emission lines. Finally, the UV absorption lines produced by these transitions are an important probe of the temperature and density along the line of sight through molecular gas.

The set of rovibrational levels within excited electronic states, however, is not complete since they only include levels that are coupled to X through dipole-allowed transitions. These excited rovibrational levels within the excited electronic states contain little of the H$_2$ population. They are important only as an intermediate step in photoexcitation of H$_2$ into excited states or the continuum of X. Dissociation energies from excited electronic states into the H(1$s$) + H($nl$) systems are taken from Sharp (1971).

---

[4] The other 1s electron is suppressed here and throughout for convenience. Doubly-excited states are not considered.

The H$_2$ molecule has ortho and para states defined by the alignment of the two nuclear spins. Within the X ground electronic state, ortho (triplet nuclear spin) states have odd rotational quantum number $J$, while para (singlet nuclear spin) have even $J$. The same rule applies for the C$^-$ and D$^-$ electronic states, while for the remainder of the electronic excited states the logic is reversed; ortho is even and para is odd. Note that there are no rotational $J$ = 0 states in the C and D electronic states, because $\Lambda$ = 1, and $J \geq \Lambda$, where $\Lambda$ is the projection of the total electronic orbital angular momentum on to the internuclear axis.

Line wavelengths are derived by differencing these level energies and correcting for the index of refraction of air for $\lambda$ > 2000Å. Comparison of these theoretical wavelengths with the observed wavelengths given in Timmermann et al. (1996) and Rosenthal et al. (2000) show that our wavelengths derived from electronic transition energies agree with the observed values to typically within $\delta\lambda/\lambda \sim 2 \times 10^{-5}$.

Spectroscopic notations, such as "2-1 S(1)", are commonly used to identify the transitions. The first numbers indicate the change in vibrational levels. The capital letter indicates the branch, which specifies the change in $J$, as shown in table 1. The number in parenthesis is the lower $J$ level of the transition. Radiative decays between ortho and para are not possible because of the different nuclear spin. Within the ground electronic state, lines have the selection rule $\Delta J = 0, \pm 2$. For electronic transitions between $\Sigma$ and $\Pi$, the selection rule is $\Delta J = 0, \pm 1$, whereas the corresponding rule for $\Sigma$ to $\Sigma$ is $\Delta J = \pm 1$. For C$^+$, the selection rule is $\Delta J = \pm 1$ whereas the corresponding rule for C$^-$ is $\Delta J = 0$. There is no such selection rule for vibrational quantum numbers.

## 2.2 Bound-bound transitions

### *2.2.1 Transitions within the ground electronic state*

Quadrupole radiative transition probabilities for rovibrational levels within the ground electronic state are taken from Wolniewicz et al. (1998). This data set is complete, i.e., transition probabilities are given for all allowed transitions between the 301 rovibrational levels.

Rate coefficient fits for collisional excitation processes within the X state are given by Le Bourlot et al. (1999; http://ccp7.dur.ac.uk/cooling_by_h2/ ) and adopted here. These include rovibrational excitation of H$_2$ by collisions with H, He, and H$_2$, but only for non-reactive transitions, i.e., those that do not involve changes in nuclear spin.

The Le Bourlot et al. (1999) compilation, obtained with quantum mechanical methods, should be appropriate to temperatures as low as ~10 K. However, this compilation contains only rate coefficients of ~500 transitions, roughly 4% of the 12,535 possible transitions within X. The data set presented by Martin & Mandy (1995; 1998) includes far more transitions but were not intended to be applied to the low temperatures needed for the PDR models as they were obtained with classical trajectory methods. Figure 4 compares their rate with the Le Bourlot et al. (1999) values for rate coefficients for collisions between H$_2$ and H for a typical transition within X. They diverge by large factors below ~400 K.

The biggest effect of the missing collision data would be to introduce errors in the populations and the heating or cooling effects of highly excited levels within X. It seems more correct to use some physically motivated rate coefficients for the missing levels rather than assume a rate coefficient of zero. We applied a version of the "g-bar" approximation (van Regemorter 1962). Figure 5 shows collision rates for all transitions within the Le Bourlot et al. (1999) data set at a given temperature, plotted as a function of the transition energy. We took the simplest possible approach, fitting these rates with the following function of energy

$$\log(k) = y_0 + a \max(\sigma, 100)^b \tag{1}$$

where $\sigma$ is the transition energy in wavenumbers and $k$ is in cm$^3$s$^{-1}$. The max function prevents unphysically large rate coefficients from occurring when the energy difference is very small. This fit is also shown in Figure 5, and Table 2 gives the fitting coefficients. These fits were within 1 dex of the majority (typically 60% to 70%) of the rates. Tests presented below will show the impact of these very approximate rates on the predicted spectrum.

We note that collisions of electronically excited states are neglected due to the negligible population in these levels as noted in Section 2.1. This is appropriate for densities well below $10^{15}$ cm$^{-3}$.

### 2.2.2 Ortho – para conversion

The ortho and para H$_2$ states can be mixed by exchange collisions with H, H$^+$, and H$_3^+$. H$^0$ and H$^+$ are the dominant forms of H at shallow depths into a PDR, while deep within the PDR, all three species have only trace abundances. Rate coefficients for H$^0$ – H$_2$ exchange collisions are from Sun & Dalgarno (1994; for $v$=0 and $J$=0-1, 0-3, and 1-2 collisions). The rate coefficients for all other H$^0$ – H$_2$ exchange collisions are taken to be zero. The approximations to the rates given in this paper diverge for temperatures below 100 K. Below this temperature a rate of zero is assumed.

Rate coefficients for H$^+$ - H$_2$ exchange collisions were taken from Gerlich (1990). The Gerlich data include both upward and downward transitions, but they do not satisfy detailed balance. We adopted the rate coefficients for the downward transitions and obtain the upward from detailed balance. These are by far the fastest collisional rate coefficients, being ~$10^{12}$ times larger than the H$^0$ rate coefficient. This is by far the dominant process for ortho-para conversion.

However, Gerlich (1990) only computed rate coefficients for $v = 0$ and $J < 10$. We therefore adopted the same procedure as discussed above to obtain the "g-bar" approximation. The Gerlich rate coefficients were fitted as a function of transition energy and used to estimate rate coefficients for all other transitions. Tests show that the estimated rate coefficients have a small effect on the final level populations and the ortho-para ratio. Le Bourlot (1991) notes that $H_3^+$ ortho-para conversion collision rates are roughly equal to the $H^+$ rates. We have used the $H^+$ rates to include collisions with $H_3^+$. $H_3^+$ generally has a small abundance and so does not strongly affect the results.

Figure 6 shows the effect of the "g-bar" approximation on $v =0$ level populations at 500K. The lower rotational level populations where collisional data exist have very little dependence on the approximation. In addition, ortho $H_2$ can be converted into para on grain surfaces (Le Bourlot 2000). This is described in the section on grain physics below.

## 2.3 Coupling to the continuum

### 2.3.1 Electronic transitions and dissociation

Transition probabilities between excited electronic states and X were taken from Abgrall et al. (2000). Excited electronic states can decay back into the bound levels within X or dissociate into the continuum of the ground state. Dissociation probabilities and the associated kinetic energy of the unbound particles were also taken from Abgrall et al. (2000).

Allison & Dalgarno (1969) give photodissociation cross sections for transitions from X into the continuum of the excited electronic states. These data are for the energy range near the threshold and do not resolve the molecule into rotation levels. The threshold for this process lies in the Lyman continuum of hydrogen for lower $v, J$ levels, and this continuum will be strongly absorbed by atomic hydrogen in a PDR. The threshold does lie within the Balmer continuum for higher $v, J$ levels, however.

This process cannot be treated with great precision because of the lack of *J*-resolved cross sections. We approximated the Allison & Dalgarno (1969) data as a single cross section, $2.5 \times 10^{-19}$ cm$^2$, and evaluated the threshold energy for each level.

### 2.3.2 Collisional dissociation from very high J

For temperatures within a PDR, $T \approx 500$ K, collisional dissociation can only occur from the very highest levels within X. Lepp & Shull (1983) give a general expression for collisions between H and H$_2$ that are valid for any *J*, but are intended for much higher shock temperatures. Hassouni et al. (2001) give rate coefficients for all *v* but only *J* = 0 for both H$_2$ + H$_2$ and H$_2$ + H. For other impactors Martin & Mandy (1995; 1998) give collisional dissociation rate coefficients, but at higher temperatures.

We have taken a broad mean of the above references to obtain the rough expression, used for collisional dissociation of H$_2$ by impact with all colliders;

$$r\left(H_2 \Rightarrow 2H\right) \approx 10^{-14} \exp\left(-\chi_{diss}(v,J)/kT\right) \text{ [cm}^3\text{ s}^{-1}\text{]} \qquad (2)$$

where $\chi_{diss(v,J)}$ is the dissociation energy. Tests show that this process has little effect on the predicted column densities and on the overall spectrum.

Electron collisions must be included as well. We know of no *J*-resolved calculations of the electron collisional dissociation rates or cross sections. Celiberto et al. (2001) give cross sections for excitation from X into B or C, followed by dissociation. This process is only fast at temperatures much higher (~50,000 K) than we expect in a PDR (~500 K). We do not include this process. Stibbe & Tennyson (1999) give rates for dissociation by electron excitation from X to the triplet state b. This is energetically far more favorable, and can occur at 500K. Unfortunately no *J*-resolved calculations exist, and dissociation from the very highest *J* states should be the fastest.

The Stibbe & Tennyson (1999) data suggest an electron rate, in the small $\chi_{diss}(v,J)/kT$ limit, that is similar to the hydrogen rate given above. We assume this rate in including electrons.

## 2.4 Grain processes

Our grain model can treat arbitrary grain types and size distributions. The built-in Mie code can resolve the grain size distribution into as many size bins as desired. The model can resolve the grain charge in multiple discrete charge states, and can also treat the stochastic heating effect. Our treatment incorporates the formalism described by Weingartner & Draine (2001). van Hoof et al. (2004) provide further details and show that our representation reproduces the Weingartner & Draine (2001) results in several test cases. In this approach the temperature and charge for each grain type and size are determined by balancing heating – cooling and ionization – recombination processes. Grain surface recombinations of electrons and ions are included in the grain charge as well as the ionization balance of the gas and the electron density. This gives a better representation of the interactions between the gas and grains since the rates are temperature dependent, and the grain temperature is set by the environment. This unified treatment is applied to PAH's (Polycyclic Aromatic Hydrocarbons) as well as larger grains.

### *2.4.1 H₂ formation on grains*

H₂ is predominantly formed on grains in the ISM although formation through the H⁻ route is not negligible. We use the rate given by Cazaux & Tielens (2002, equation 18)

$$r_g = \sum_{n_g} 0.5 \times n(H^0)\left[n_g \sigma_g / n(H)\right] u_H \, \varepsilon(H_2) \, S_H(T) \quad [\text{s}^{-1}] . \tag{3}$$

Here the sum is over all grain types and sizes and the factor of 0.5 accounts for the fact that two hydrogen atoms become a single H₂ molecule (see equation 3a of Biham et al. 1998). The total hydrogen density *n(H)* is given by $n(H) = n(H^0) + n(H^+) + \sum_{other} n(HX)$, where HX represents all other H-bearing molecules, and is basically used as a surrogate for the total grain density within the term in the square brackets. This entire term is the projected grain area per unit hydrogen nuclei. The atomic density *n(H⁰)* multiplies this term rather than *n(H)* so that the rate becomes small in the limit where the gas is fully molecular and little *H⁰* is present (see equation 1 of Hirashita et al. 2003). The remaining variables are $u_H$, the average speed of a hydrogen atom, $\varepsilon(H_2)$, the H → H$_2$ conversion efficiency, and $S_H(T)$, the H atom sticking probability. We take the dimensionless sticking probability from Hollenbach & McKee (1979);

$$S_H(T) = \frac{1}{1 + 0.4(T_2 + T_{g2})^{0.5} + 0.2 T_2 + 0.08 T_2^2}, \tag{4}$$

where $T_2$ is the gas kinetic temperature in units of 100 K and $T_{g2}$ is the grain temperature in these units. We assume the H$_2$ conversion efficiency given by Cazaux & Tielens (2002; equation 16), assuming 10$^{-10}$ monolayers s$^{-1}$. Tests show that this results in a H$_2$ formation rate within 10% of the Jura (1974) rate for grain temperatures near 100 K.

Newly created H$_2$ populates various rovibrational levels within the ground electronic state. Population distribution functions have been considered by Black & van Dishoeck (1987), Le Bourlot et al. (1995), Draine & Bertoldi (1996) and more recently by Takahashi (2001) and Takahashi & Uehara (2001). We follow the last two references. The other two distributions are also included in the code as additional options to allow tests for their impact on results. We use the Takahashi & Uehara (2001) vibrational distribution, and take the average of the two forms of the rotational distribution, equations 6 and 7 in Takahashi (2001). We assume that no population occurs when the $E_{rot}$ function of Takahashi & Uehara (2001), their equation (3), becomes negative, as happens for large *v*. This distribution function depends on the grain composition. For reference, the carbonaceous, silicate, and ice compositions tend to deposit newly created $H_2$ in the $v$=2, 6, and 7 levels, respectively. We also used their product energy distribution of newly formed $H_2$ on various dust grains as given in Takahashi & Uehara (2001).

### *2.4.2 Ortho – para conversion on grain surfaces*

Conversion from ortho to para $H_2$ can occur on grain surfaces at low temperatures. We follow Le Bourlot (2000) in treating this process. The rate is given by

$$r_g = \sum_{n_g} \left( n(H_2) \middle/ n(H) \right) n(H) \left( n_g \sigma_g / n(H) \right) u_{H_2} \, \eta(T_{ad}, T_d) \, S_{H_2}(T) \quad [\text{s}^{-1}] \,. \tag{5}$$

We assume that the sticking probability $S_{H_2}$ for molecular hydrogen is equal to that for atomic hydrogen, and use the Hollenbach & McKee (1979) probability. In implementing this we took a binding energy of $T_{ad}$ = 800 K. A temperature $T_{crit}$ is defined from $T_{ad}$. This critical temperature is of the order of 20 K for these materials.

The ortho to para conversion efficiency is then given by $\eta_c(T_{ad}, T_d)$. We use the form given in Le Bourlot (2000). In this process $H_2$ in any $v$, $J$, level strikes the grain and is deexcited during its time on the surface. When $T_{grain}$ > $T_{crit}$ the $H_2$ is deexcited to either $J$=0 or 1, preserving the initial nuclear spin, before leaving the grain surface. When $T_{grain}$ < $T_{crit}$ all $H_2$ go to $J$=0. Since $T_{ad}$ is so low, the ortho-para conversion will only be efficient in deep, well shielded, parts of a PDR.

## 2.5 The chemistry network

The chemistry network in the code has been described by Ferland et al. (1994; 2002). We have reviewed and updated all aspects of the chemistry to the data sets given by Abel et al. (1997), Galli & Palla (1998), Hollenbach et al. (1991), Maloney et al. (1998), Le Teuff et al. (2000), and Stancil et al. (1998). Only details that affect the abundance and level populations of $H_2$ are described here.

The chemistry network includes ~1000 reactions including 66 species involving hydrogen, helium, carbon, nitrogen, oxygen, silicon, and sulphur. Tests made during the Leiden PDR workshop (http://hera.ph1.uni-koeln.de/~roellig/) show good agreement between our calculations and other PDR models. Time-dependent advective terms are included when dynamical models are considered (Ferland et al. 2002), and $H_2$ may not be in a time-steady state in some PDRs (Bertoldi & Draine 1996). The simulations presented here are all stationary and represent gas that has had time to reach statistical equilibrium.

The micro-physical model of $H_2$, described in Sections 2.1-2.4, is not explicitly part of this network. Rather total formation and destruction rates are used to determine a total $H_2$ density. Following Tielens & Hollenbach (1985; hereafter TH85), the molecule is divided into ground state $H_2^g$ and excited state, $H_2^*$, and these two are part of the $H_2$ and CO chemistry networks. The density of $H_2^*$ is the sum of all populations in states with an excitation potential greater than 2.6 eV, and $H_2^g$ as the population in lower levels. This was chosen to follow the partitioning given by TH85, which then allows us to use their chemical rate coefficients for reactions that involve $H_2^*$. The CO photodissociation rate is calculated as in Hollenbach et al. (1991). The detail calculations concerning our chemistry network will be in future publication.

The equilibrium deduced from the network is then used to set the state specific $H_2$ formation and destruction rates that affect the population with X. In turn the micro-physical model of the hydrogen molecule defines state specific destruction rates, which are sums over the state-specific rates. Destruction rates predicted by the micro-physical $H_2$ model are used in the chemical reaction network, and state specific creation rates from the network are then included as population mechanisms within the large molecule. The net effect is that each level has a population that is determined by these creation and destruction processes, along with internal excitations and deexcitations within the molecule. The solutions obtained by the micro-physical model of the hydrogen molecule (see Section 2.7 for a discussion of how the level populations are calculated) and the full chemistry network are converged by multiple solution of both; so that the H$_2$ density from the chemistry network and the sum of populations within X agree to a specified amount.

### 2.5.1 Creation from H$^-$

Although H$_2$ is mainly formed on grain surfaces in the ISM, it can also form by the associative detachment reaction

$$H^- + H \rightarrow H_2 + e^-. \tag{6}$$

Our detailed treatment of this process is described in Ferland & Persson (1989). This is usually the fastest in moderately ionized regions (Ferland et al. 1994). Abel et al. (1997) note that H$_2$ is mostly formed in the para configuration ($v=0$, $J=0$ and $v=0$, $J=2$) by this process.

We use the state-specific distribution functions for H$_2$ formation from H$^-$ given by Launay et al. (1991), and populate the newly formed H$_2$ molecule using these rates and the H$^-$ density determined by the molecular network. The population distribution function tends to peak around $v \sim 5$ to 7 and mostly odd $J$, typically between 3 and 7, although very large $J$ can occur.

The Launay et al. distribution function does not, at face value, agree with the Abel et al. (1997) distribution functions. Abel et al. (1997) assume that after formation the molecule will reside only in low $J$ levels ($J = 0$-2). According to Abel et al. (1997), the rate coefficient for the process H + e$^-$ → H$^-$ + γ is smaller than the rate coefficient for the process H$_2$ ($J=2$) + H$^+$ → H$_2$ ($J=1$) + H$^+$ + 170.5 K (given by Flower and Watt 1984 and used by Abel et al. 1997), so the molecule decays to lower levels more quickly than it is formed. So, newly formed H$_2$ will eventually be converted to the para state ($v=0$, $J=0$ and $v=0$, $J=2$). Abel et al. (1997) give the final state after the system has had time to relax, which is different from the state-specific rates given by Launay et al. 1991 (Ortho-para conversion is treated separately.)

The chemistry network uses an integrated rate for the formation of H$_2$. We summed over all of the state specific rates in Launay et al. (1991) and fitted the total rate with the following:

$$k(H^-) = \frac{1}{5.4597 \times 10^8 + 71239\, T} \ [\text{cm}^3\ \text{s}^{-1}]\ . \qquad (7)$$

This expression is well-behaved for all temperatures, but was fitted to the Launay et al. (1991) data over the temperature range they give, 10 K – 10$^4$ K.

Similarly, the process H$_2^+$ + H → H$_2$ + H$^+$ produces H$_2$ in $v$=4 states. This process can be important in deeper, molecular regions of a PDR. We use rates from Krstić (2002). These are not $J$-resolved so we place all in $J$ = 0.

## 2.6 High-energy effects

Cosmic rays can be included in a calculation as described by Ferland & Mushotzky (1984) and Wolfire et al. (1995). X-rays have very nearly the same effect. Although X-rays have little effect upon a gas at shallow depths, since relatively few high-energy photons are present, deep within a cloud the attenuated radiation field can be dominated by hard photons. These predominantly interact with inner shells of the heavy elements producing hard photoelectrons and inner shell vacancies which relax, producing high-energy Auger electrons. Our treatment of this physics is described in Ferland et al. (1998). All of these processes create electrons with a great deal of energy, and within neutral or molecular regions, these energetic electrons create secondary ionizations and excitations, while depositing only part of their initial energy as heat.

The effects of non-thermal electrons in molecular gas are described by Dalgarno et al. (1999), and in the papers on the chemistry of the ISM given above. In particular, Tiné et al. (1997) give cosmic ray excitation rates for $v$=0 to 14 and $J$=0-11 within X which we adopt. We adopt their values at a fractional ionization $10^{-3}$ as being typical.  Rates for electronic excitation are from Dalgarno et al (1999).

## 2.7 Determining level populations

The molecule is represented by the 1893 levels shown in Figure 1.  Level populations are determined by solving a system of simultaneous linear equations.  The level populations are evaluated several dozens of times per radial step as the pressure, temperature, and ionization are simultaneously determined. Conventional matrix inversion methods have a speed that goes as a power of the number of levels, and would result in prohibitive execution times. Additionally, many matrix inversion methods have stability problems with very large matrices when the level populations range over several decades of orders of magnitude. Rather, a linearized method was developed which takes into account the physical properties of the molecule.

### 2.7.1 Excited electronic levels

Collisional excitation to electronic states is completely negligible due to the low temperature in a PDR. Excitation is by photoexcitation in discrete lines, the Solomon process, which can be followed by relaxation back into rovibrational levels of the ground electronic state or by transitions into its continuum.  The electronic excited states have short lifetimes because of their large transition probabilities.  We neglect collisional processes within excited electronic states since the critical densities, the density where the collisional deexcitation and spontaneous radiative decay rates are equal, is ~$10^{14}$ cm$^{-3}$, far greater than the densities encountered in a PDR.

The line photoexcitation rate is given by $A_{u,l} \eta_\nu g_u / g_l$ [s$^{-1}$], where $\eta_\nu = J_\nu / (2h\nu^3/c^2)$ is the photon occupation number of the attenuated incident continuum at the line frequency (Ferland & Rees 1988) and self absorption of a single line is treated as in Ferland (1992).  Line overlap is discussed below.  The excited level decays by allowed radiative transitions. The population of an electronic excited state $n_u$ is then given by the balance equation

$$\sum_l n_l \, \eta_\nu \, \frac{g_u}{g_l} A_{ul} = n_u \left[ A_c + \sum_l A_{ul} (\beta + \delta) \right] \tag{8}$$

where $A_c$ indicates the transition probability from the excited state into the X continuum, $A_{ul}$ is the transition probability between X and the excited states, $\beta$ is the escape probability and $\delta$ is the destruction probability. We follow the treatment described in Ferland & Rees (1988). The total rate for $H_2$ to decay into the X continuum is then given by

$$r_c = \frac{1}{n(H_2)} \sum_u n_u \, A_c \;\; [\text{s}^{-1}], \tag{9}$$

which is the total destruction rate due to the Solomon process.

### 2.7.2 Populations within ground electronic state

Highly excited states within X are mainly populated by indirect processes. The most important include decays into X from excited electronic states and formation in excited states by the grain or H- route. Direct collisional excitation is relatively inefficient, at least for higher $J$, due to the large excitation energies and low gas kinetic temperatures.

We use a linearized numerical scheme that updates populations by "sweeping" through the levels within X, from highest to lowest energies. Rates into and out of a level are computed and an updated level population derived from these rates. At the end of the sweep the updated and old populations are compared, and revisions continue until the largest change in population is below a threshold. The method achieves a solution in a time that is proportional to $an^2$, where $n$ is the number of levels and $a$~5 is the number of iterations needed to determine the $H_2$ populations. By comparison to standard linear algebra, solving the system with LU decomposition would require a time $\approx \frac{1}{3} n^3$, or $10^3 - 10^4$ times slower.

This process converges quickly because of the complete linearization that is inherent to the entire code. During solution of the ionization, chemical, and thermal equilibria within a particular zone, the $H_2$ molecule populations will be determined several dozens of times, and each new solution will be for conditions that are not too far from previous solutions.

Two convergence criteria were used. The first is that the level populations have stabilized to within a certain tolerance, usually 1%. This affects the observed emission-line spectrum. Collisional deexcitation of excited levels within X can be the dominant gas heating mechanism in parts of a PDR (TH85), and a second criterion was that the net heating (or cooling) have stabilized to better than a small fraction of the heating-cooling error tolerance. These criteria were chosen so that the molecular data remained the main uncertainty in the solution.

### 2.7.3 Line overlap

For an isolated line we use equations A8 and A9 of Federman et al. (1979) to treat coarse line overlap of the Doppler core and damping wings. For the case of the electronic lines of $H_2$, many thousands of lines and the heavy element ionizing continua overlap. Neufeld (1990), Draine & Bertoldi (1996), and Draine & Hao (2002) discuss ways of treating overlapping lines. Cloudy has always included the effects of line overlap for important cases, which were added on an individual basis (one case is discussed by Netzer et al. 1985). A general solution to the overlap problem, with overlap treated in an automated manner is necessary. We developed a method that looks to the long term evolution of the code, since the overlap problem will only become worse as the simulations become more complete and more lines are added.

The simplest approach is to make the continuum energy mesh fine enough to resolve all lines; then lay the line opacities and source functions onto this mesh. This approach is taken, for instance, by the Phoenix stellar atmosphere code for atomic lines (Hauschildt & Baron 1999). It is not feasible to do this in Cloudy since the code's execution time is linearly dependent on the number of continuum points. This is because of the repeated evaluation of photo-interaction rates and continuous opacities – the far higher spectral resolution needed to resolve lines would result in prohibitive execution times.

We use a multi-grid approach to treat line and continuum overlap. In this scheme certain aspects of a problem are solved on a fine mesh, with far higher spectral resolution, which is then propagated out onto a courser energy scale. The coarse continuum cells are many hundreds of line-widths across, so it is not possible to tell, on that scale, whether lines actually interact and to what extent. The fine mesh has several fine-resolution continuum elements per linewidth; the total line opacities and source functions are placed onto this fine mesh, and the effects of line overlap is handled trivially. Lines are individually transferred within the fine mesh and then referenced back onto the course grid to include interactions between line and the continuum. An example of the coarse and fine continua is presented in Section 3.3 below.

## 2.8 Heating – cooling

Processes that add kinetic energy to the free particles are counted as heating agents, while cooling processes remove kinetic energy from the gas. $H_2$ affects the thermal state of its surroundings in several ways.

The dissociation from electronic states following photoexcitation is an important heating process, which is given by a modified form of equation 9,

$$G_c = \sum_u n_u A_c \varepsilon_c \text{ [erg cm}^{-3}\text{ s}^{-1}] \tag{10}$$

where the sum is over all electronic excited states, $A_c$ is the dissociation probability, and $\varepsilon_c$ is the kinetic energy per dissociation, as tabulated by Abgrall et al. (2000).

Collisional processes are included for all levels within $X$, as described above. The energy exchange associated with these collisions, expressed as a net cooling, is given by

$$L_X = \sum_u \sum_{l<u} (n_l C_{lu} - n_u C_{ul}) \varepsilon_{lu} \quad [\text{erg cm}^{-3} \text{ s}^{-1}] \tag{11}$$

where the sum is over all levels within $X$, and the $C$'s represent the sum of all collisional rates that couple levels $u$ and $l$. $\varepsilon_{lu}$ is the difference in energies between the levels. For conditions close to the illuminated face of a PDR, higher levels are overpopulated relative to a Boltzmann distribution at the gas kinetic temperature and collisions within $X$ is a net heating process. Deep within the cloud the process acts as a coolant.

## 3 Comparison calculations

This section describes several comparison calculations that illustrate aspects of the $H_2$ simulations. We begin with the simplest case, a model in which the molecule is dominated by thermal collisions. Next the case of cosmic ray excitation is considered, then a situation similar to the illuminated face of a PDR, in which populations are determined mainly by the Solomon process. Finally a complete calculation of a standard PDR is presented, with the resulting $H_2$ emission spectrum and column densities. These cases have been well established by previous studies, for instance Abgrall et al. (1992), Black & van Dishoeck (1987) and Sternberg & Neufeld (1999), but our main purpose is to document the current implementation of the relevant physics. If no special values are mentioned then all the model parameters have the values listed in Table 3.

## 3.1 Collision dominated regions

Collisions play an important role in determining $H_2$ spectra in shocks, or gas that is otherwise heated to warm temperatures. As a test case we have computed the collisionally excited spectrum of a predominantly molecular gas.

In the low density limit, where every collisional excitation is followed by emission of a photon, the collisional rate is given by $\langle \sigma u \rangle n(H_2) n_c$, where the first term is the excitation rate coefficient and $n_c$ is the density of colliders. The emissivity is then given by $4\pi j = \langle \sigma u \rangle n(H_2)^2$ [erg cm$^{-3}$ s$^{-1}$]. In this limit the ratio $4\pi j / n(H_2)^2$ should depend mainly on the collision physics.

Figure 7 shows the spectra of collisionally dominated gas at 500K, a typical PDR temperature. Two hydrogen densities, 10$^3$ cm$^{-3}$ and 10$^5$ cm$^{-3}$, are shown. Figure 8 shows collisionally dominated spectra at 2000K, a typical shock temperature, with the same densities. The Solomon process and cosmic ray excitation were disabled to make the excitation due entirely to collisions with thermal particles. The gas was almost entirely molecular ($n(H_2)/n(H) = 0.5$). Line emissivities are greater for higher temperature as expected. For both models the long wavelength transitions, which form from smaller *J* in *v* = 0, have a higher emissivity for the lower density. These lines are collisionally deexcited at the high density. As the density increases, the higher ro-vibrational levels become more populated by collisions from intermediate *J* resulting in more intense lines originating from those high levels.

## 3.2 Cosmic ray dominated regions

Both very hard photons and cosmic rays may penetrate into regions where hydrogen is predominantly molecular. In such environments the primary effect of collisions between thermal matter and energetic particles or light is to produce non-thermal electrons which then excite or ionize the gas. This can result in both destruction and excitation of $H_2$. We assume the galactic background ionization rate quoted in Williams et al. (1998), $\zeta/n_H$ = 2.5×10$^{-17}$ cm$^3$ s$^{-1}$, where $\zeta$ is the primary cosmic ray H$^0$ ionization rate (s$^{-1}$), and the excitation distribution function of Tiné et al. (1997), as described above. This cosmic ray background will be included in many of the models described below since this background ionization process should be pervasive.

Non-thermal electrons can also excite rovibrational levels of H$_2$, as described above. As a test, we recomputed a 50 K model. The low temperature was chosen so that cosmic ray excitation dominates. The electronic lines were forced to be optically thick to make the Solomon excitation process unimportant, and as a result the gas is almost entirely molecular. The predicted spectrum is shown in Figure 9. Inspired by Tine et al. (1997), we also predicted a spectrum at 50K. The differences are because of the full chemistry of this model, likely to be different, which affects the thermal excitation rates. Given this uncertainty the agreement with their calculation is quite satisfactory.

The cosmic ray dominated spectra is much richer in character than the collision dominated spectra described above. The energy levels of H$_2$ are widely separated so most molecules reside in the ground vibrational level for fully molecular environments. Thermal collisions cannot excite high rovibrational levels. Cosmic rays are highly energetic and can pump H$_2$ in higher rovibrational levels resulting in numerous intense lines.

### 3.3 The Orion PDR

We consider the TH85 standard model since this has parameters chosen to be similar to inner regions of the Orion complex and the model has become a standard within the PDR community. Table 3 summarizes the parameters of the TH85 Orion PDR. We assume the constant gas density, chemical composition, and incident radiation field given in TH85. Ferland et al. (1994) discuss a previous comparison of our results with this paper.

As outlined in section 2.4 above, our treatment of the grain physics has improved, and this is a major source of differences with our earlier calculation. We use a grain size distribution that is chosen to reproduce the overall extinction properties of the extinction along the line of sight to the Orion environment. Most of this extinction occurs within the Veil, a predominantly atomic layer of gas that lies several parsecs towards the observer from the central star cluster. The column density per unit extinction is taken from Shuping & Snow (1997) and Abel et al. (2004). We have also included PAHs in our calculation. Grain charge, temperature, and drift velocity are self-consistently determined for each type and size of grain.

As a starting point we consider a point near the illuminated face of the PDR. Figure 10 shows the populations per molecule for an unshielded atomic gas with a constant temperature of 500 K. Collisions are relatively unimportant and the $H_2$ level populations are determined by the Solomon process. As expected, the populations are highly non-thermal, with large populations in high $J$ and $v$ levels

Next we compute the full ionization, chemical, and thermal structure of the PDR. Figures 11 and 12 show the computed abundances of $C^+$, C, CO, $H^+$, H, and $H_2$. For $A_V$ > 2.4, $H_2$ is the dominant species while all C is in the form of CO for $A_V$ > 4.4. Figure 13 shows the computed temperature structure and the electron density. At the illuminated face of the cloud the electron temperature is > 200 K and $T$ drops to 30K deep inside the cloud. At the edge of the cloud the electron abundance is determined by the ionization of C and is equal to the $C^+$ abundance. These results are in qualitative agreement with the original TH85 calculation.

Collisional de-excitation of $H_2$ is a significant gas heating mechanism. TH85 present a widely used theory for the process that involves a simple representation of the $H_2$ molecule. They use two levels, $H_2$ and $H_2^*$, and their heating rate (equation A13) is linearly dependent on $H_2^*$. Conversely, we consider all 301 bound levels in the ground electronic state. Figure 14 compares the heating due to the $H_2$ de-excitation between TH85 and our model. Deep inside the cloud, the $H_2^*$ population is very small, since it is predominantly produced by the Solomon process, which is inefficient due to line self-shielding. This small population makes the TH85 de-excitation heating rate very small. In our calculation heating at depth is mainly due to collisions with low $J$ levels, which are not included in the $H_2$ and $H_2^*$ representation. As a result, deep inside the cloud these two are considerably different.

Photoelectric emission from grains is a major source of heating. As discussed by Ferland et al. (1994), our model of grain photoelectric heating results in generally lower temperatures than the original TH85 calculation. As discussed above, our current treatment follows Weingartner & Draine (2001), and results in slightly less heating. Figure 15 shows the variation of temperature for the graphite and silicate grains and PAH. The PAHs are hotter than the other two due to their smaller sizes, while the graphitic component is generally hotter than the silicate. Within each type of grain, smaller grains are hotter than larger due their reduced radiation efficiency. The dependence on grain size and composition is in the expected sense. The $H_2$ grain formation rate depends on grain temperature, which shows the importance of using a realistic representation of the grain physics in this simulation.

The Solomon process is the most important destructive mechanism for $H_2$. Figure 16 compares the Solomon process dissociation rates given by equation 9 with that computed using the approximation given by equation A8 of Tielens & Hollenbach (1985) and by equation 23 given by Bertoldi & Draine (1996).
At the illuminated face of the cloud our model predicts greater Solomon rates than the other two. This higher value of the Solomon rate has been observed by the other groups present at the Leiden PDR workshop (2004). The effect is due to the inclusion of 301 levels in $X$, which will produce more pumping from the excited rovibrational levels to the higher electronic states resulting in a larger Solomon rate. Deep inside the cloud self-shielding plays a crucial role and our model matches better with the Draine & Bertoldi model (1996).

The integrated column densities for various $J$ levels in the $v = 0$ level are shown in Figure 17(a) by filled circles, whereas the LTE column densities are given by open circles. It is clear from the plot that the $J$ = 0-3 levels are in LTE. The populations of these low levels are mostly affected by collisions and henceforth have come into equilibrium. The populations of higher levels are mainly populated by nonthermal pumping processes and are overpopulated relative to their LTE value. The gas temperature can be derived safely for these conditions from the ratio of column densities of $J$ = 1 to $J$ = 0 since these levels are in LTE. We find a mean gas temperature of 49.4K from the ratios of column densities shown in Figure 17(a). Ratios of higher rotational levels ($J > 3$) can be used as diagnostic indicators of nonthermal pumping. The integrated column densities for various $J$ levels for the $v$ = 1-8 vibration levels are shown in Figures 17(b) and 17(c). This type of diagnostic diagram is commonly used to determine conditions within translucent molecular regions (Draine & Bertoldi, 1996). Most of the $H_2$ is in the ground vibrational level and its column density is much larger than the excited vibrational levels. The column density decreases with increasing $v$. The populations, and their variation with $J$, of the excited vibrational states are different than the ground vibrational state and the distinction between ortho and para states are more prominent for $v >0$.

The full $H_2$ spectrum is shown in Figure 18. $H_2$ can offer additional PDR diagnostic indicators of the density and flux of the ionizing photons. Note that optical lines should be detectable, although at very faint levels.

We also calculated the total ortho to total para ratio as a function of $A_V$ as given by Abgrall et al. (1992) and Sternberg & Neufeld (1999). This is shown in Figure 19. This has a similar characteristic as the previous calculations.

Earlier we discussed the treatment of line overlap. Figures 20(a) and 20(b) show the coarse and fine continua for the point at the half-molecular location in the PDR, where $2n(H_2)/n(H) = 0.5$. Figure 20(a) shows the coarse continuum, used in the evaluation of the continuous opacities and photodissociation rates. Thermal emission by dust dominated the infrared emission, with fine structure atomic emission superimposed. PAH features are present in the near IR, and a set of weak $H_2$ lines are present in the 1-10 μm region. Figure 20(b) shows the fine continuum transmission factors. This is similar in appearance to, for instance, Figure 2 of Abgrall et al. (1992). We can see finer details as we zoom out the wavelength scale.

## 3.4 The Leiden (2004) PDR

We, along with eight other groups, participated in a PDR workshop held in Leiden in the Spring of 2004. The purpose was to compare predictions of various PDR codes. Three other codes included complete models of the $H_2$ molecule, and as test case was devised to check their predictions. The other codes were those described by Sternberg & Neufeld (1999), Black et al. (1987), and Le Bourlot et al. (1995). Here we compare the predictions of one of the constant temperature models (F1). The parameters are given in Table 4. For simplicity an equipartition distribution function, in which 1/3 of the energy of formation of $H_2$ goes into the grains, internal excitation, and kinetic energy, was assumed in all cases. Figure 21 compares the fractional populations at the illuminated face of the cloud as a function of excitation energy. Figure 22 shows the integrated column density for the same model as a function of excitation energy. Our results agree well with the other groups with the differences due to implementation details.

# 4 Conclusions

The aim of the current manuscript was to document the physical, chemical, and thermal processes relevant to $H_2$ and the implementation of these processes into the spectra simulation code Cloudy. Great effort has been made to treat all of these processes self-consistently in a synergistic approach. By specifying the gas composition, density, column density, dynamical state, and the incident radiation field of a particular astrophysical environment, one can uniquely determine its physical conditions and emitted spectrum. This work lays the foundation for future investigations of the role of $H_2$ in various astrophysical environments. This version of Cloudy is available at http://www.nublado.org.

# 5 Acknowledgements

We wish to thank M. Mihalache-Leca, H. Abgrall and E. Roueff, Jacques Le Bourlot, Junko Takahashi and Lutoslaw Wolniewicz for their help. Ewine van Dishoeck, Amiel Sternberg, Jacques Le Bourlot, and Evelyne Roueff graciously allowed us to present their results in our Figures 21 and 22. We would like to thank the anonymous referee for thoughtful suggestions which helped us to improve our paper. GS would like to thank CCS, University of Kentucky for their two years of support. NASA (NAG5-12020) and NSF (Ast 03-07720) also supported this work. PCS acknowledges support from NSF grant AST 00-87172. P.v.H is being supported by the engineering and Physical Sciences Research Council of the United Kingdom.

# 6 Appendix – effects of uncertain collision rate coefficients

The uncertainties in the collisional rate coefficients are substantial. If the errors are random and not correlated then we can simulate the uncertainty with

Gaussian random numbers. We assume the following form for the error distribution

$$\log r' = \log r + \text{rand}$$

where rand is a Gaussian random number with a mean of 0 and a dispersion of 0.5.

This random noise in collisional rate coefficient changes the final spectra (Figure 23) by less than 5% for very low rotational quantum numbers such as 0-0 S(0), 0-0 S(1) and 0-0 S(2), but by as much as 25% for higher rotational quantum numbers such as 0-0 S(3), 0-0 S(4). Figure 24 plots the level populations for $v$=0 with and without the random Gaussian noise in collision data at 500K.

Table 1. Branch notation

| Branch | $J_{up} - J_{lo}$ |
|--------|-------------------|
| O      | -2                |
| P      | -1                |
| Q      | 0                 |
| R      | +1                |
| S      | +2                |

Table 2. g-bar fitting coefficients

| collider    | $y\_0$   | $a$     | $b$    |
|-------------|----------|---------|--------|
| H           | -9.9265  | -0.1048 | 0.456  |
| He          | -8.281   | -0.1303 | 0.4931 |
| $H_2$(ortho)| -10.0357 | -0.0243 | 0.67   |
| $H_2$(para) | -8.6213  | -0.1004 | 0.5291 |
| $H^+$       | -9.2719  | -0.0001 | 1.0391 |

Table 3.  Model parameters for Orion PDR

| | |
|---|---|
| $n_H$(cm$^{-3}$) | $2.3\times10^5$ |
| $G_0$ | $10^5$ |
| $\delta v_d$(km s$^{-1}$) | 2.7 |
| grains | Orion, PAH |
| $\zeta_H$ (s$^{-1}$) | $2.5\times10^{-17}$ |
| Composition | See text |
| Continuum | See text |
| A$v$/$N$(H) | $4\times10^{-22}$ |

Table 4.  Model parameters for F1 model of Leiden PDR workshop (2004)

| | |
|---|---|
| $n_H$(cm$^{-3}$) | $10^3$ |
| $G_0$ (Draine field) | 10 |
| $\delta v_d$(km s$^{-1}$) | 1 |
| grains | ISM 1.16 |
| $\zeta_H$ (s$^{-1}$) | $2.5\times10^{-17}$ |
| grain temperature | 20 K |
| constant gas temperature | 50 K |

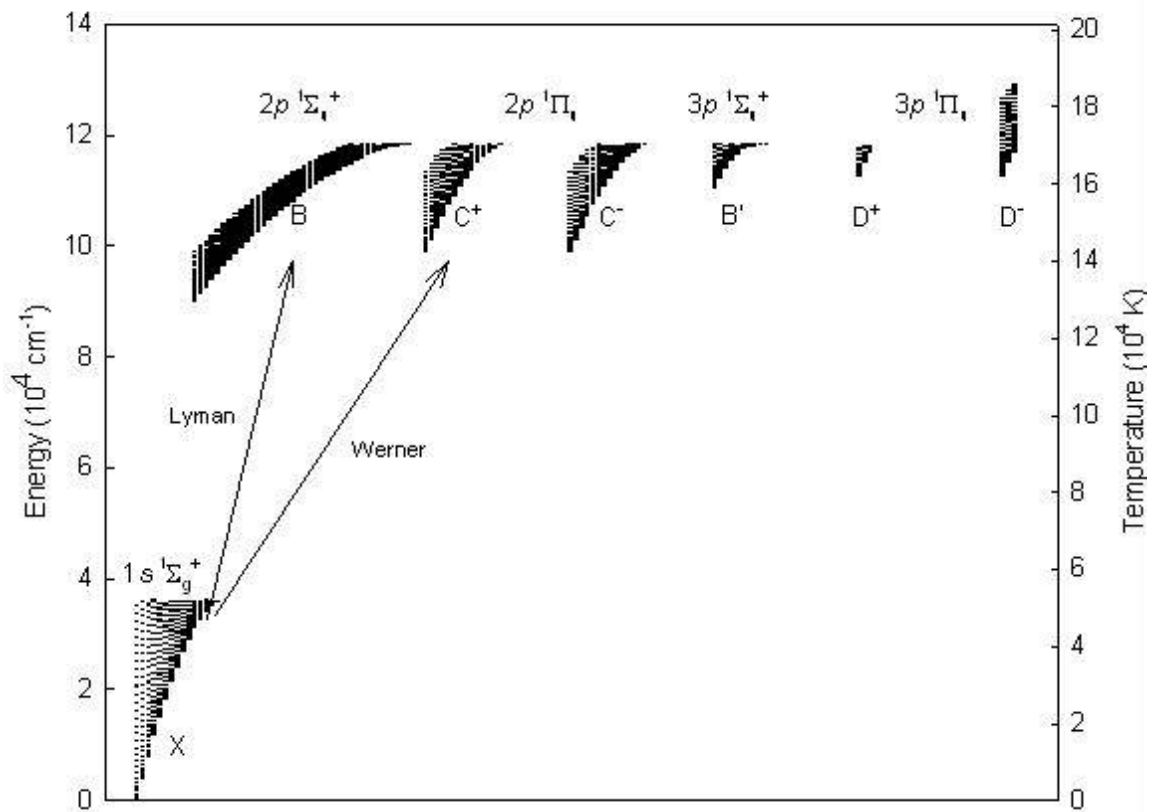

Figure 1. The full set of energy levels for all electronic states included in our calculation. The configuration (with the other 1s electron suppressed for convenience) is given above each group of levels, and the shorthand notation for the level is below.

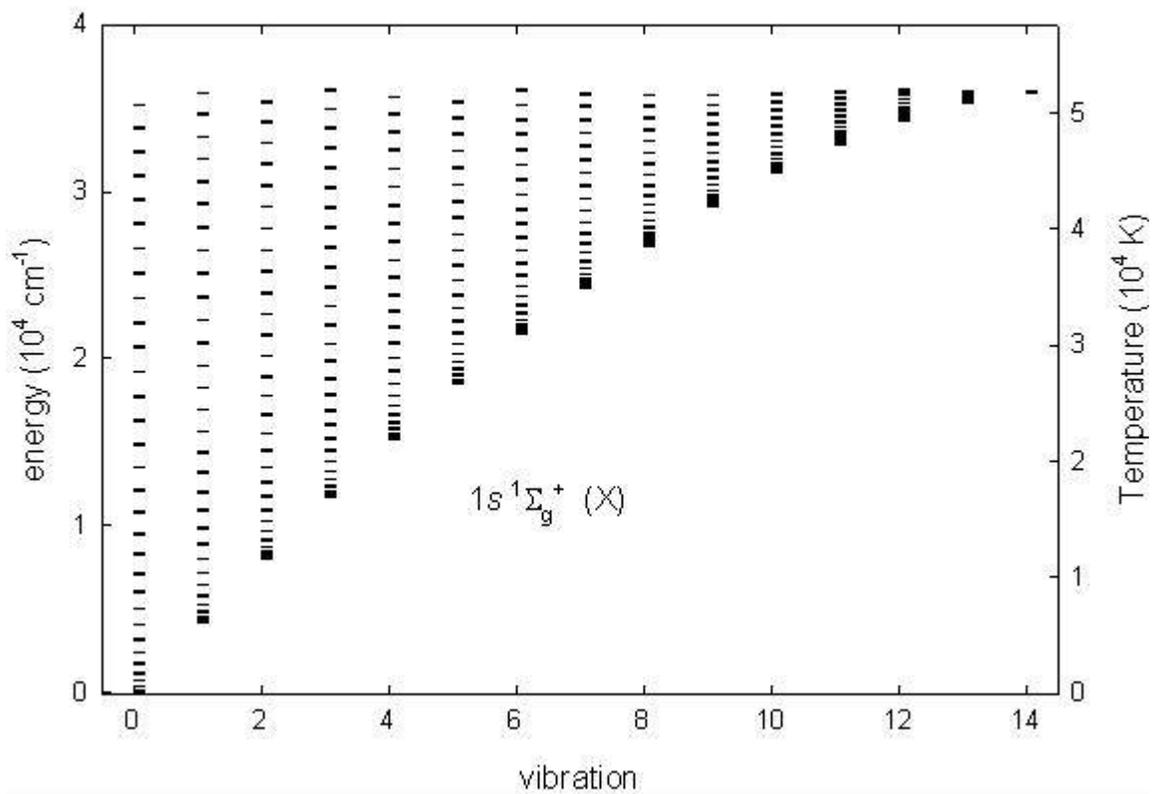

Figure 2. The energy levels within the ground electronic state. The excitation energies are given in cm$^{-1}$ relative to the lowest level ($v$ = 0, $J$ = 0). The equivalent temperature is given on the right hand axis.

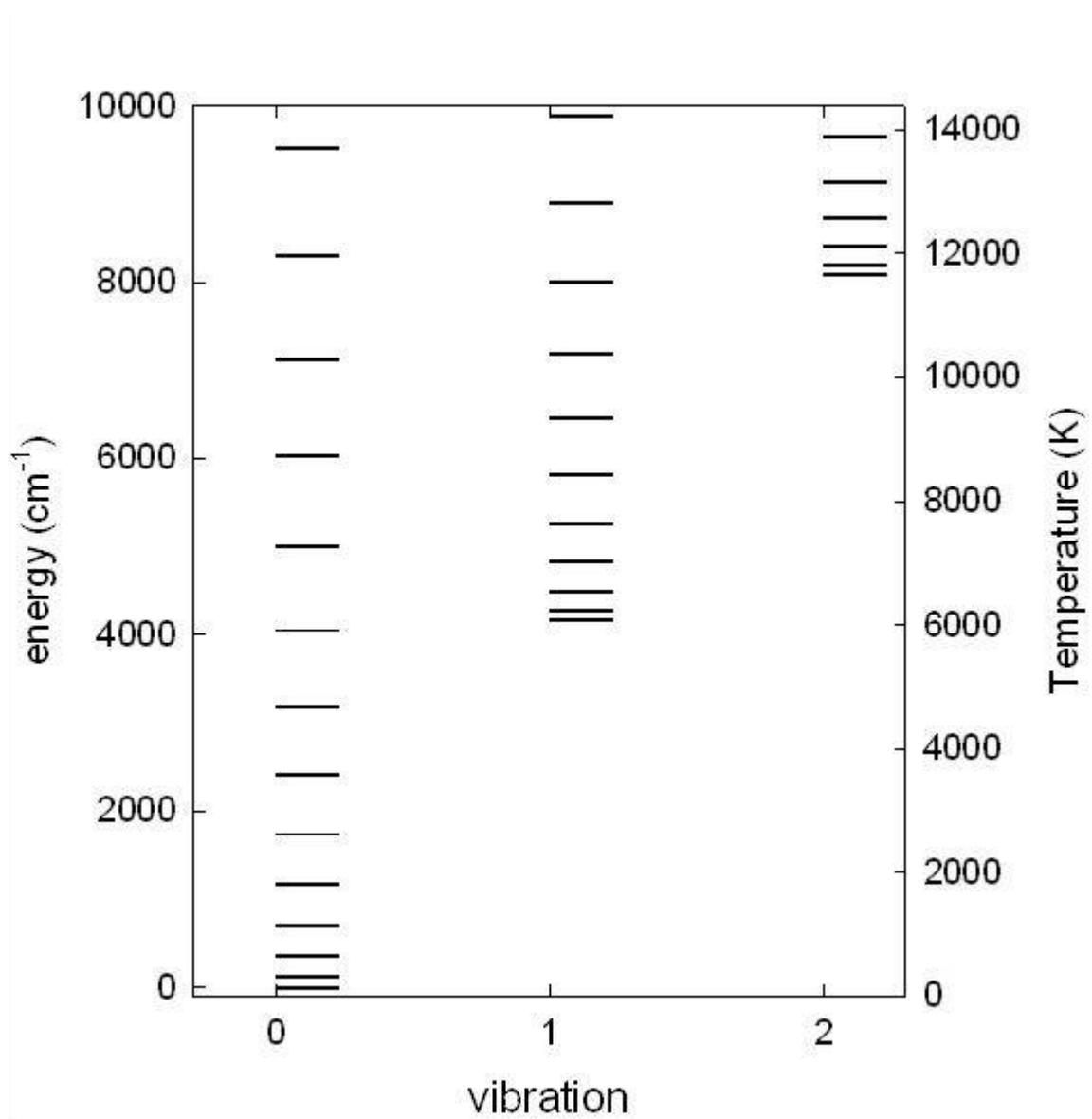

Figure 3. The lowest rotational levels within the first three vibration levels of the ground electronic state. The energy scales are in both wavenumbers and K.

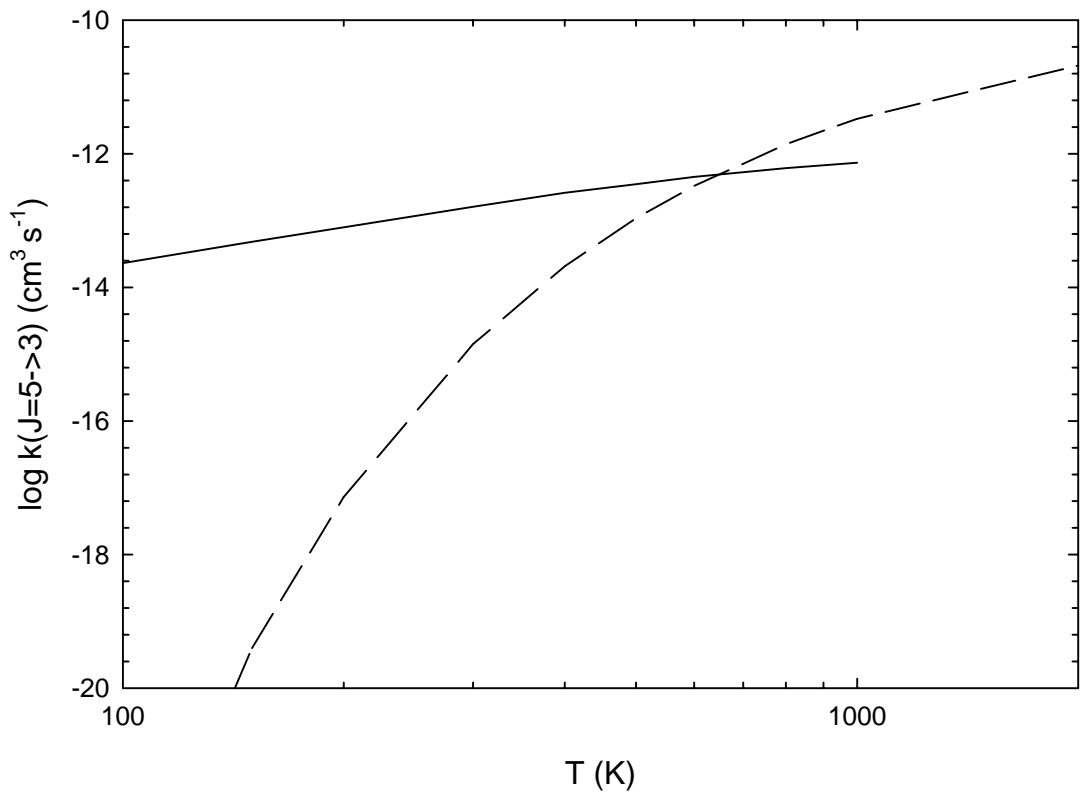

Figure 4. A comparison between the collision rate coefficients of Martin & Mandy (1995; 1998; the dashed line) and Le Bourlot et al. (1999; the solid line) for H + $H_2(v=0; J=5)$ => H + $H_2(v=0; J=3)$. They differ by large amounts for PDR temperatures (T < 400 K).

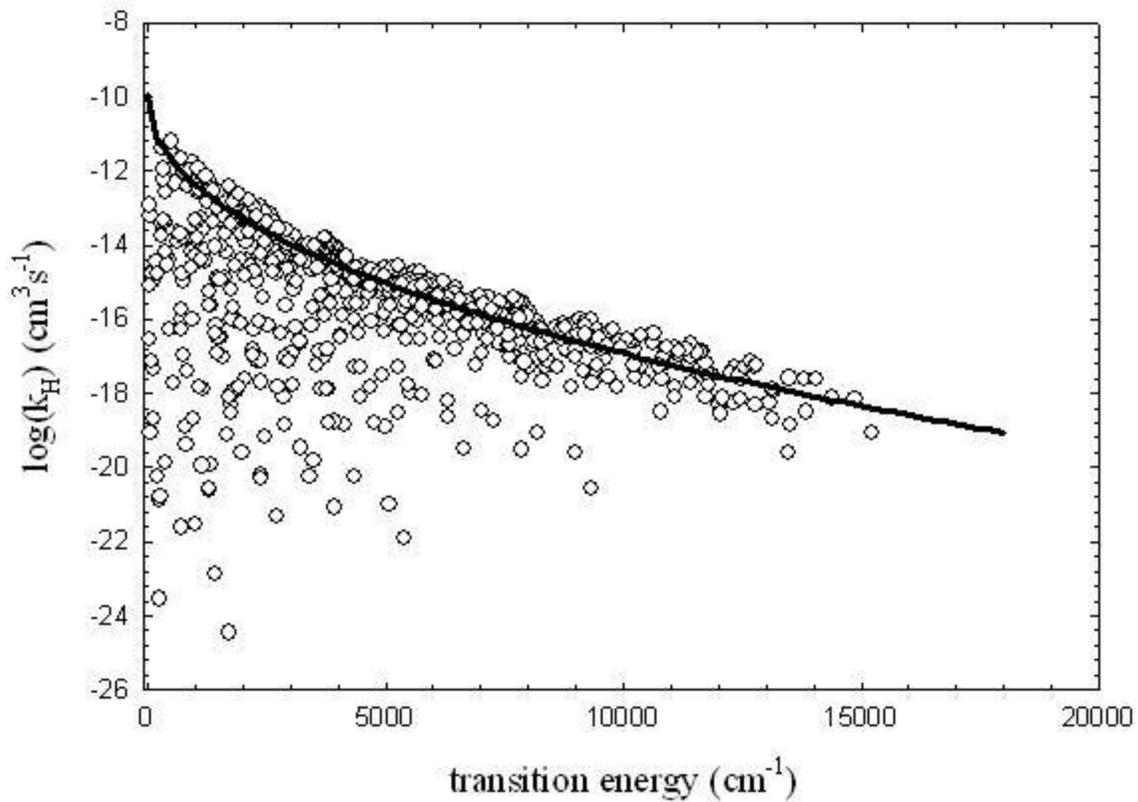

Figure 5. The collision rates as a function of transition energy for the lowest 52 transitions within the H – $H_2$ collision data of Le Bourlot et al (1999). The solid line shows a best fit to the main envelope of points. More than 50% of the rate coefficients lie with 1 dex of the line.

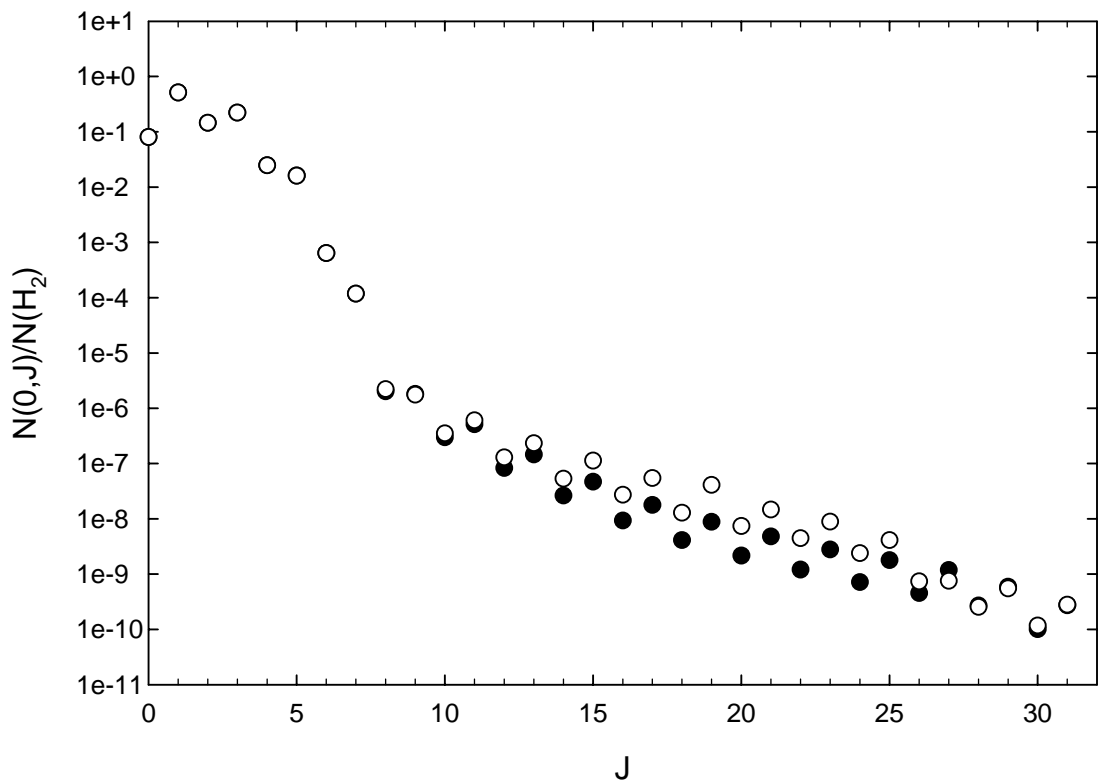

Figure 6. The level populations for *v*=0 and *T*=500 K with and without the g-bar collisional data. The rotation quantum number *J* is indicated along the x-axis while the y-axis gives the populations per $H_2$. The open circles are with the "g-bar" approximation and the black circles are without. The high *J* levels are effectively uncoupled radiatively from the other levels within the X state because most of the $H_2$ population is in low *J* levels. Further pumping into the excited electronic states results mostly in decays back into other low-*J* levels. The populations of low *J* levels, where collision data exist, have very little dependence on the "g-bar" approximation.

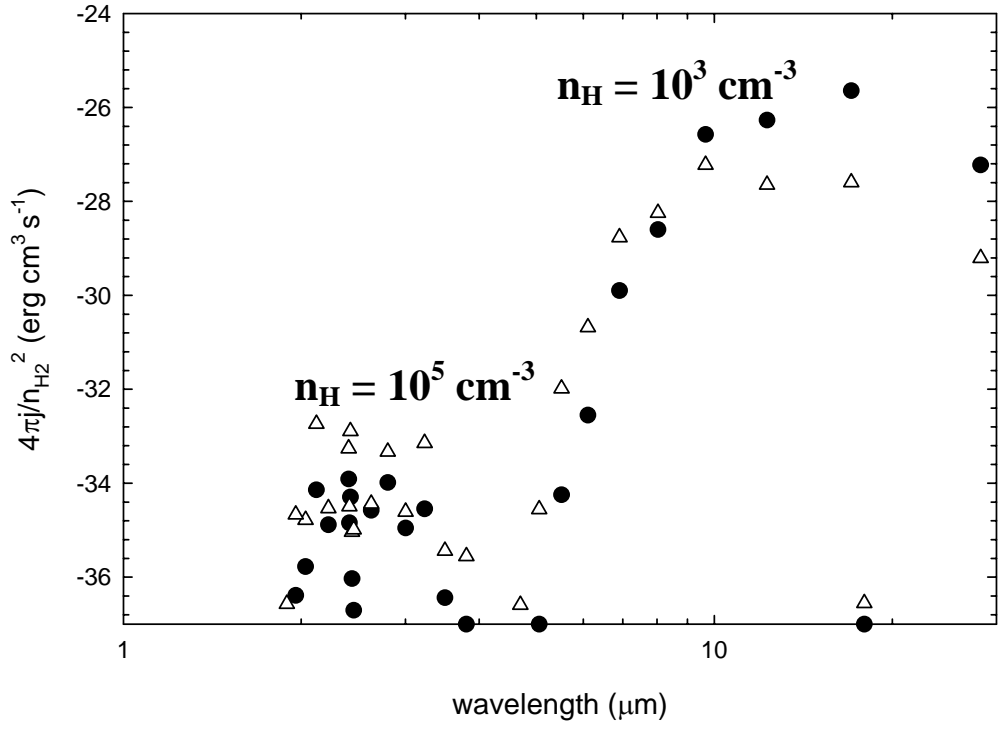

Figure 7. A spectrum of the collisionally excited $H_2$ gas at 500K with hydrogen density $10^3$ cm$^{-3}$ (filled circle) and $10^5$ cm$^{-3}$ (open triangle).

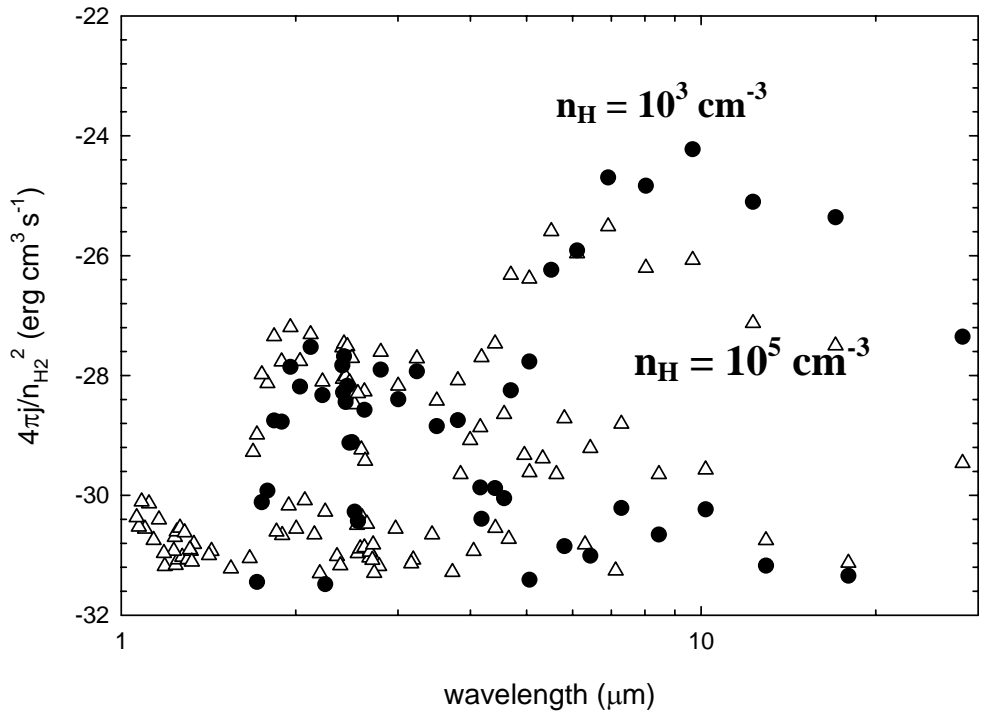

Figure 8. A spectrum of the shock excited $H_2$ gas at 2000K with hydrogen density $10^3$ cm$^{-3}$ (filled circle) and $10^5$ cm$^{-3}$ (open triangle).

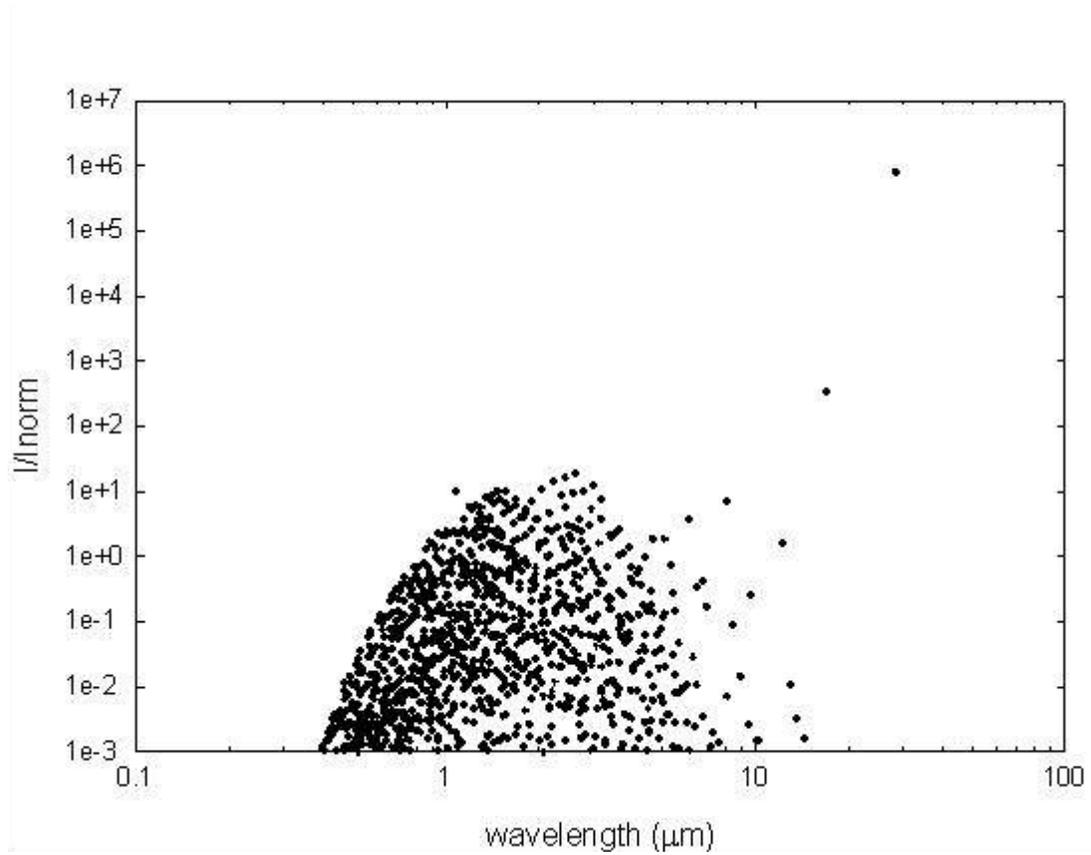

Figure 9. A spectrum of cosmic ray excited H$_2$ gas at 50K. The line intensities are predicted relative to 2.121 μm 1-0 S(1) line.

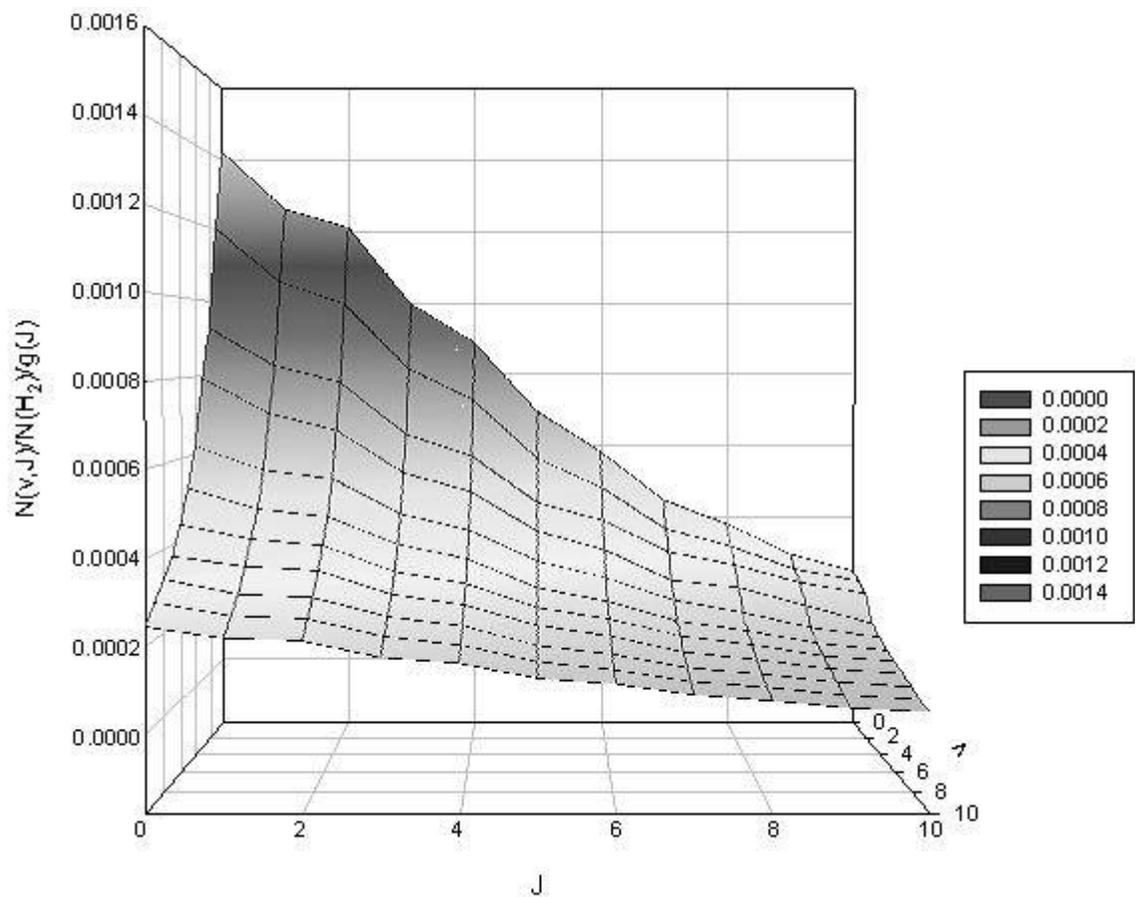

Figure 10. The populations per molecule for a 500 K gas. The calculations are for unshielded gas where the electronic lines are optically thin.

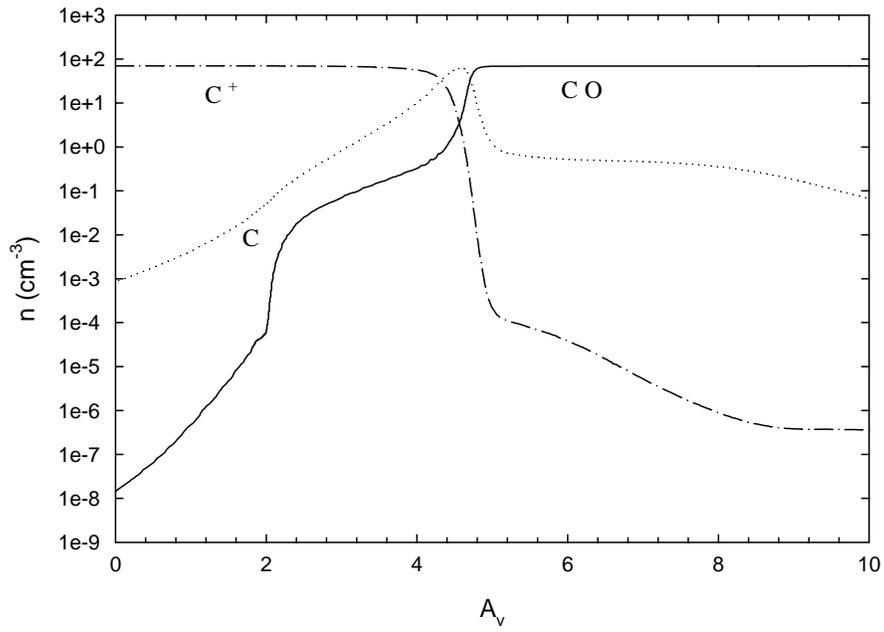

Figure 11. C (dotted line), C+ (dash-dot line), and CO (solid line) density for the standard TH85 PDR.

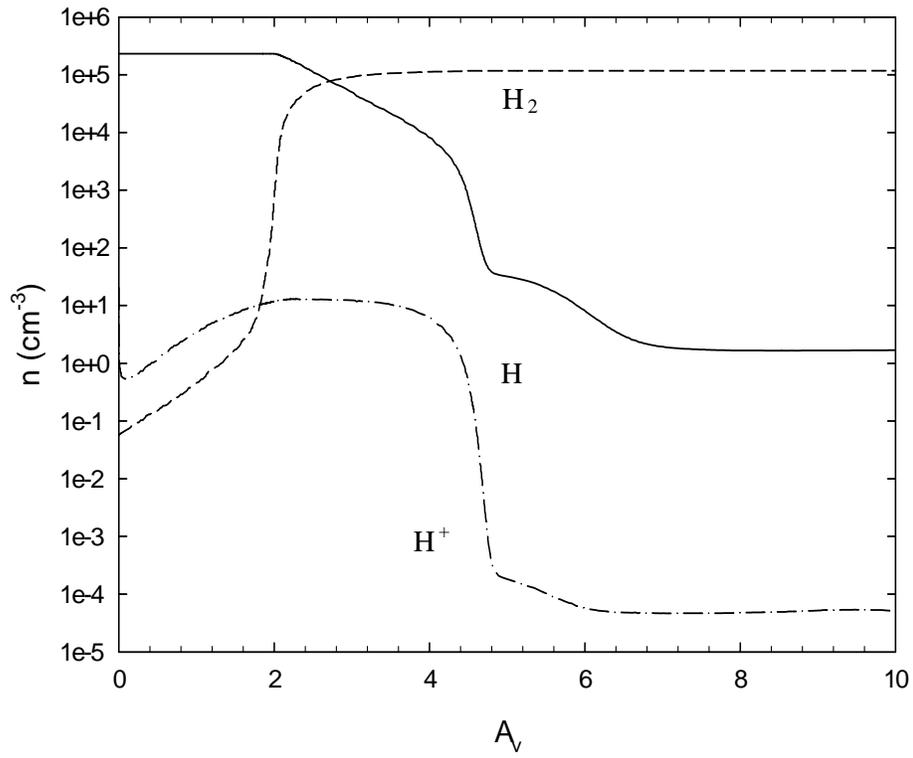

Figure 12. $H_2$(dashed line), $H^+$ (dash-dot line), and H(solid line) density for the standard TH85 PDR.

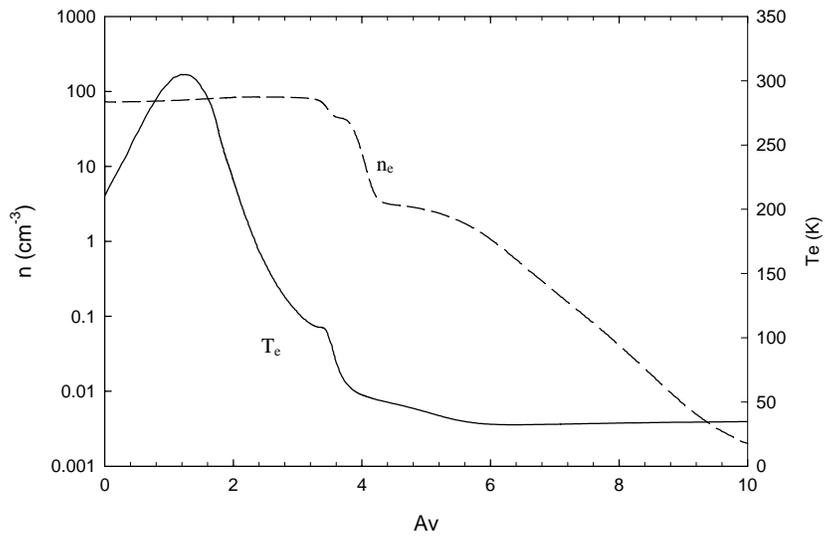

Figure 13. The electron temperature (solid line) and electron density (dashed line) from the standard TH85 PDR.

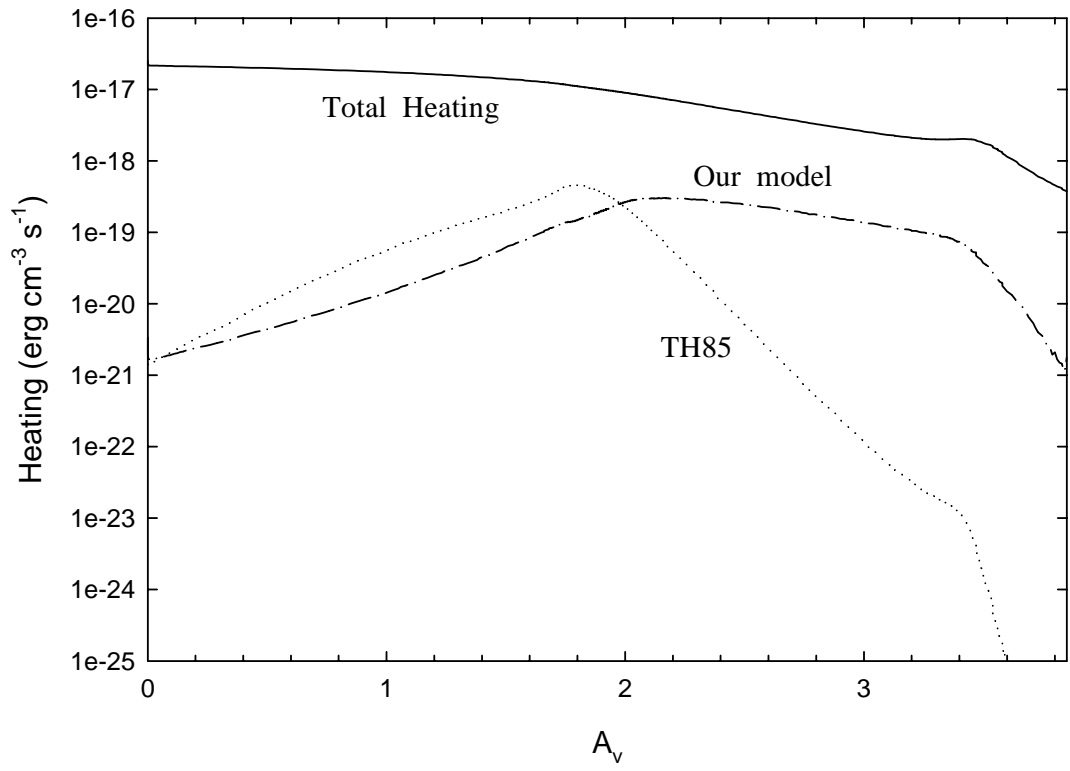

Figure 14. A comparison of heating due to $H_2$ de-excitation between TH85 and our model.

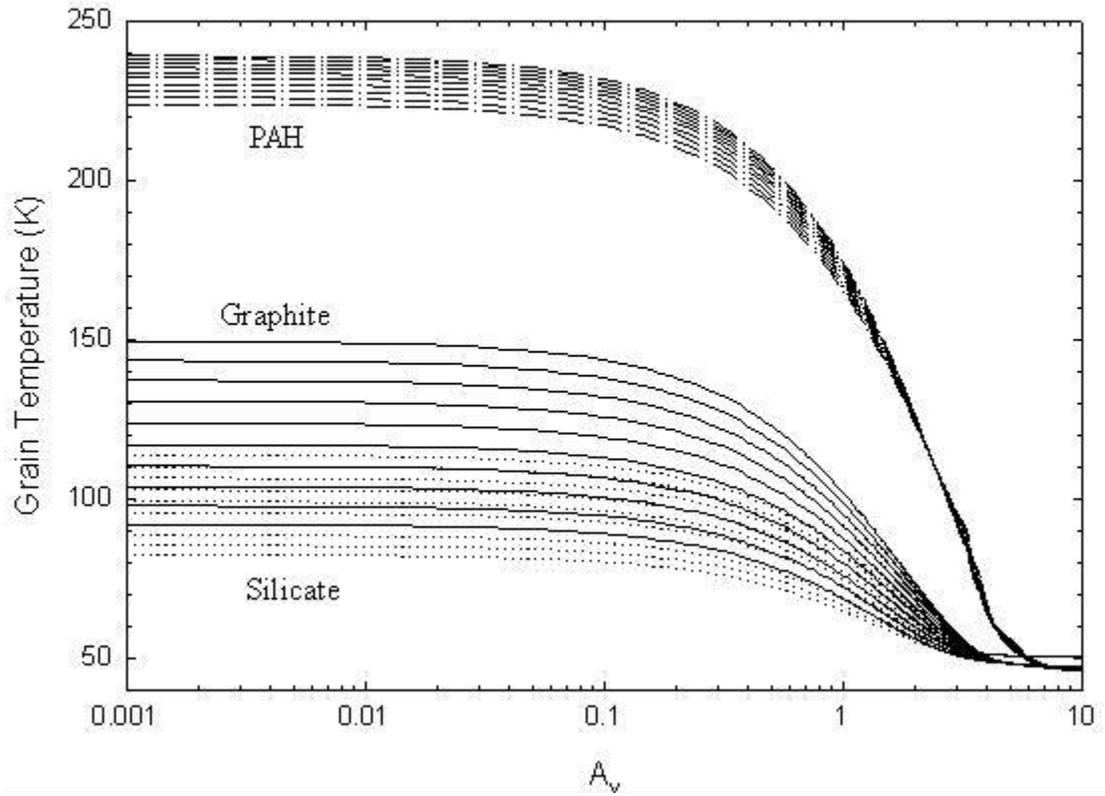

Figure 15. The grain (distributed in 10 bin sizes) temperature as a function of $A_v$.

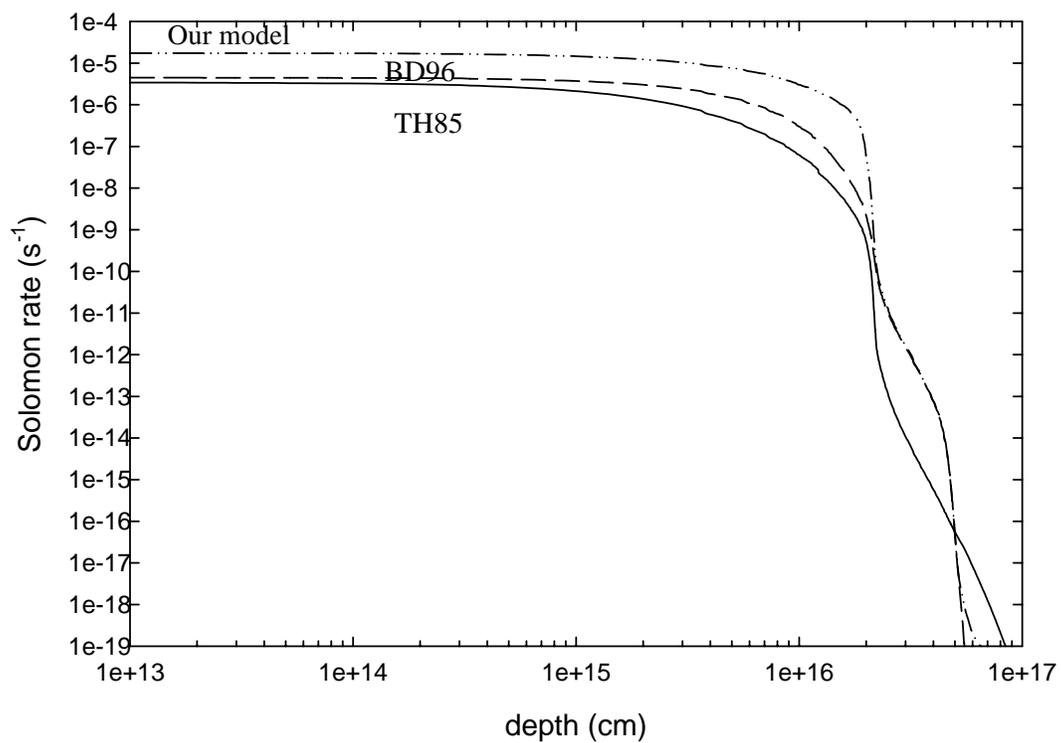

Figure 16. A comparison of several simple Solomon rates with Solomon rates from the current micro-physical model of the $H_2$ molecule. The Solid, dashed and dash-dot-dot lines represent TH85, BD96 and our model respectively.

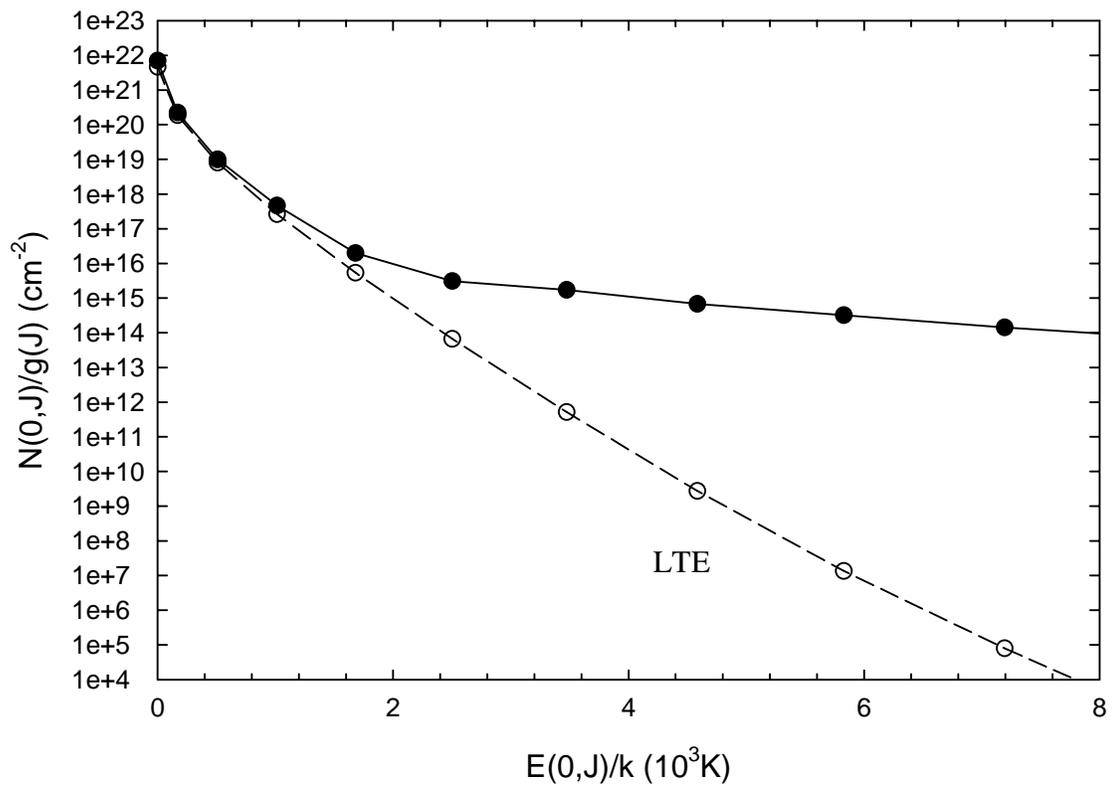

Figure 17(a). The Predicted column densities (filled circles), divided by statistical weight, for the TH85 PDR. The open circles represent column densities, divided by statistical weight, for LTE. These lower levels are in collisional equilibrium and so show no ortho-para distinction.

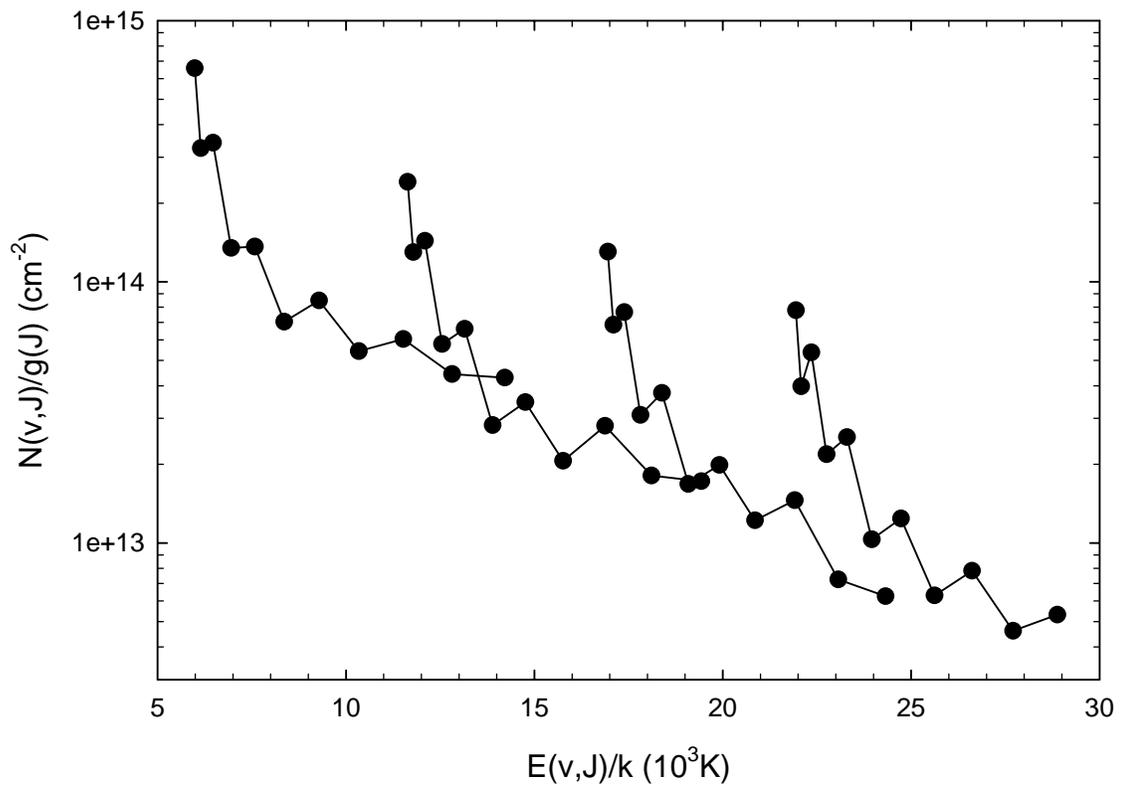

Figure 17(b). The Predicted column densities (divided by statistical weight, for *v*=1-4) for the TH85 PDR. These more highly excited levels are not in thermal equilibrium and do show differences between the ortho and para states.

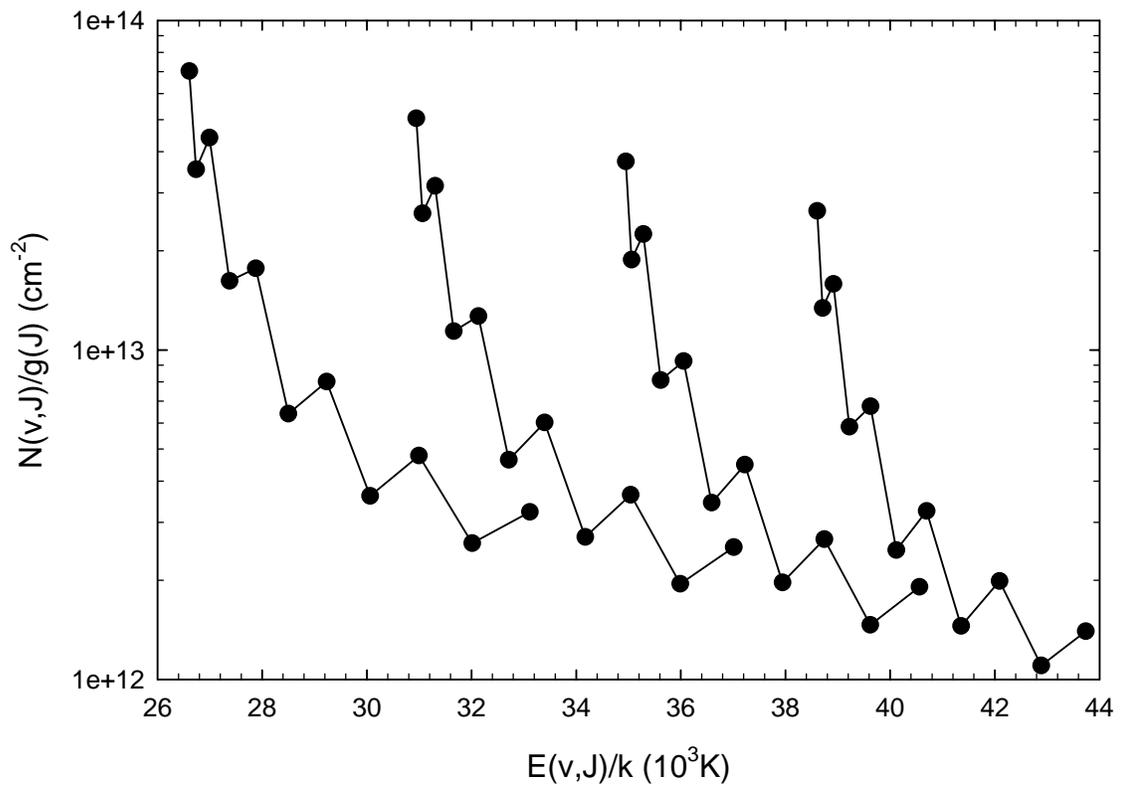

Figure 17(c). The Predicted column densities (divided by statistical weight, for *v*=5-8) for the TH85 PDR.

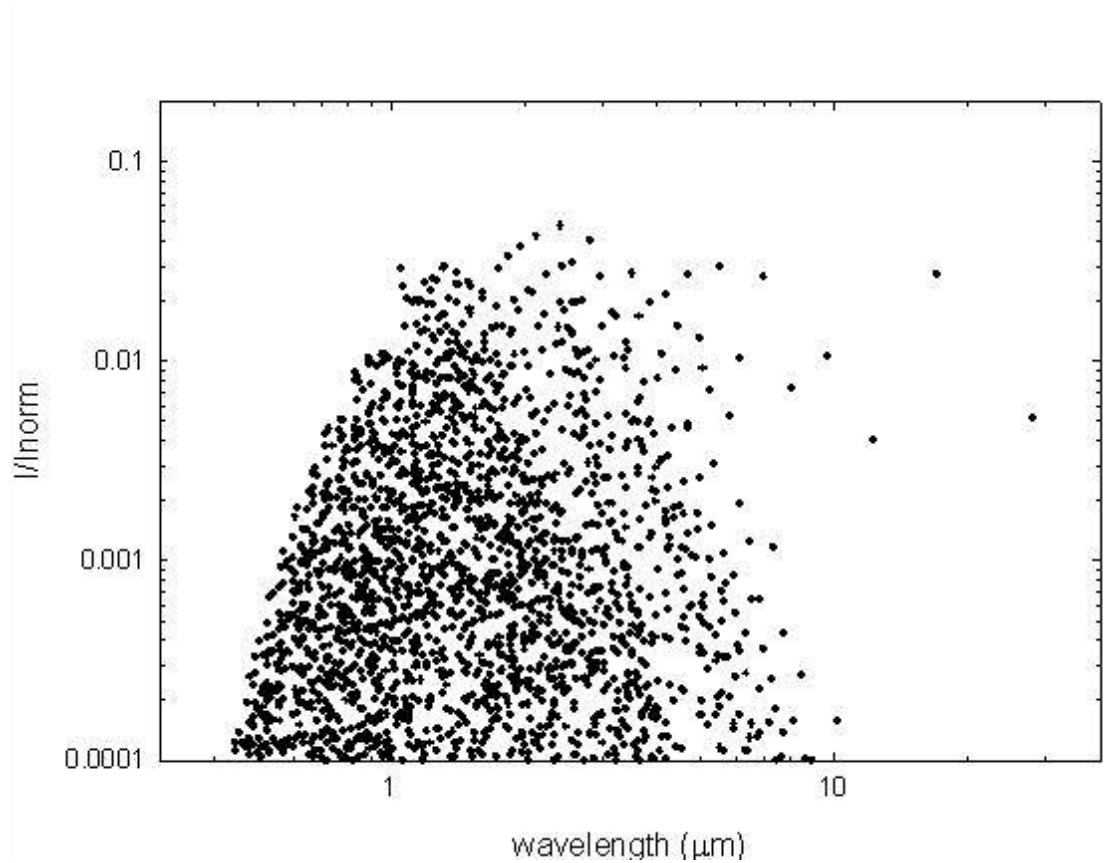

Figure 18. The emergent $H_2$ spectrum for the standard TH85 PDR. Due to the density of lines in the 0.5 – 40 μm region each emission line is plotted as a point rather than a true emission line. The line intensities are predicted relative to [C II] 157.6 μm line.

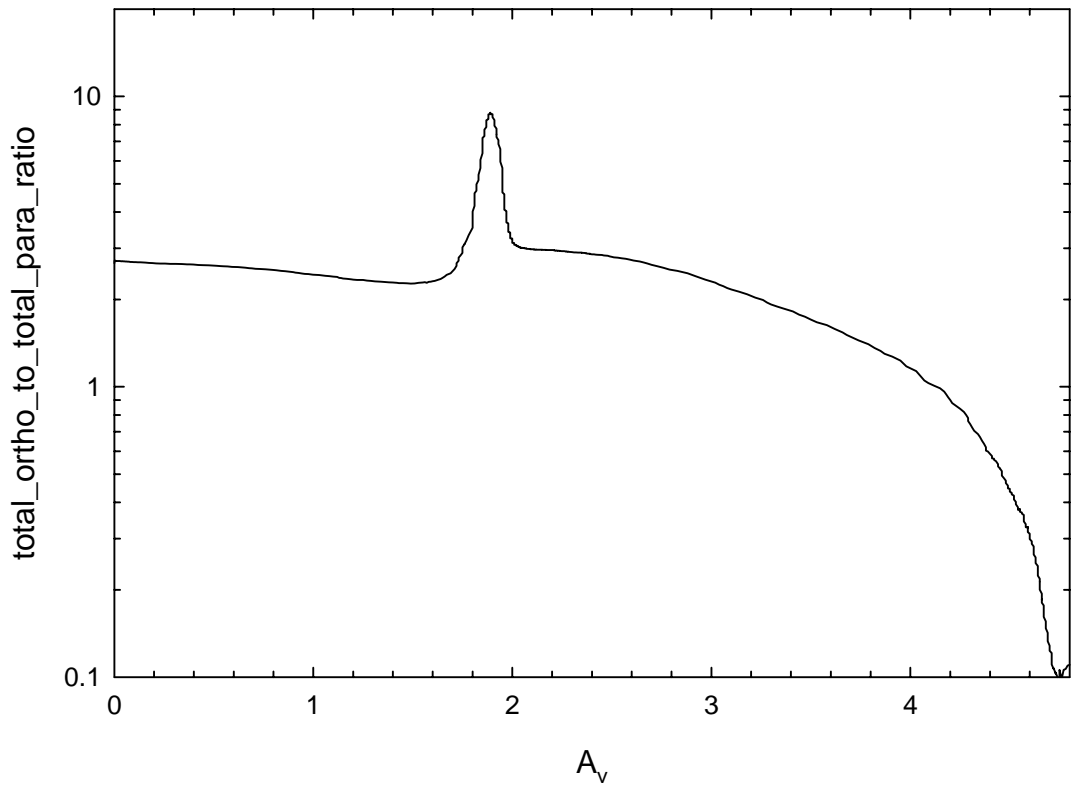

Figure 19. Ratio of total ortho to total para populations for the standard TH85 PDR.

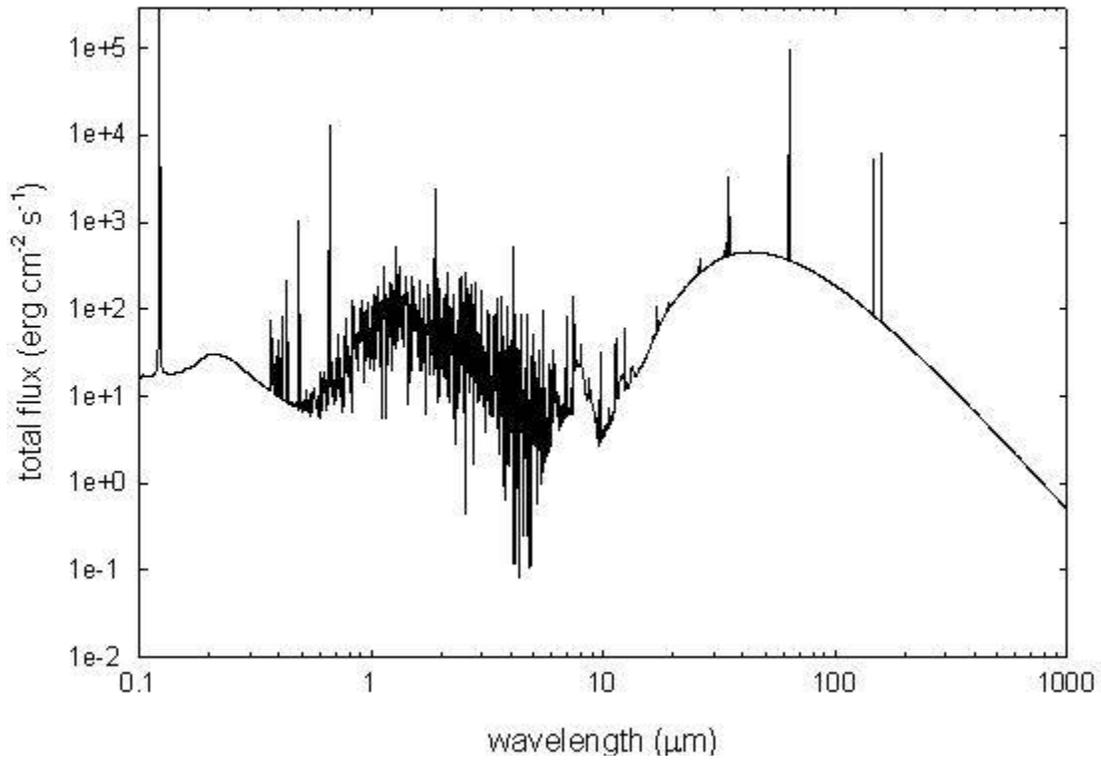

Figure 20(a). Coarse line and continuum for the TH85 PDR. This calculation stopped at the point where the hydrogen molecular fraction reached 0.5.

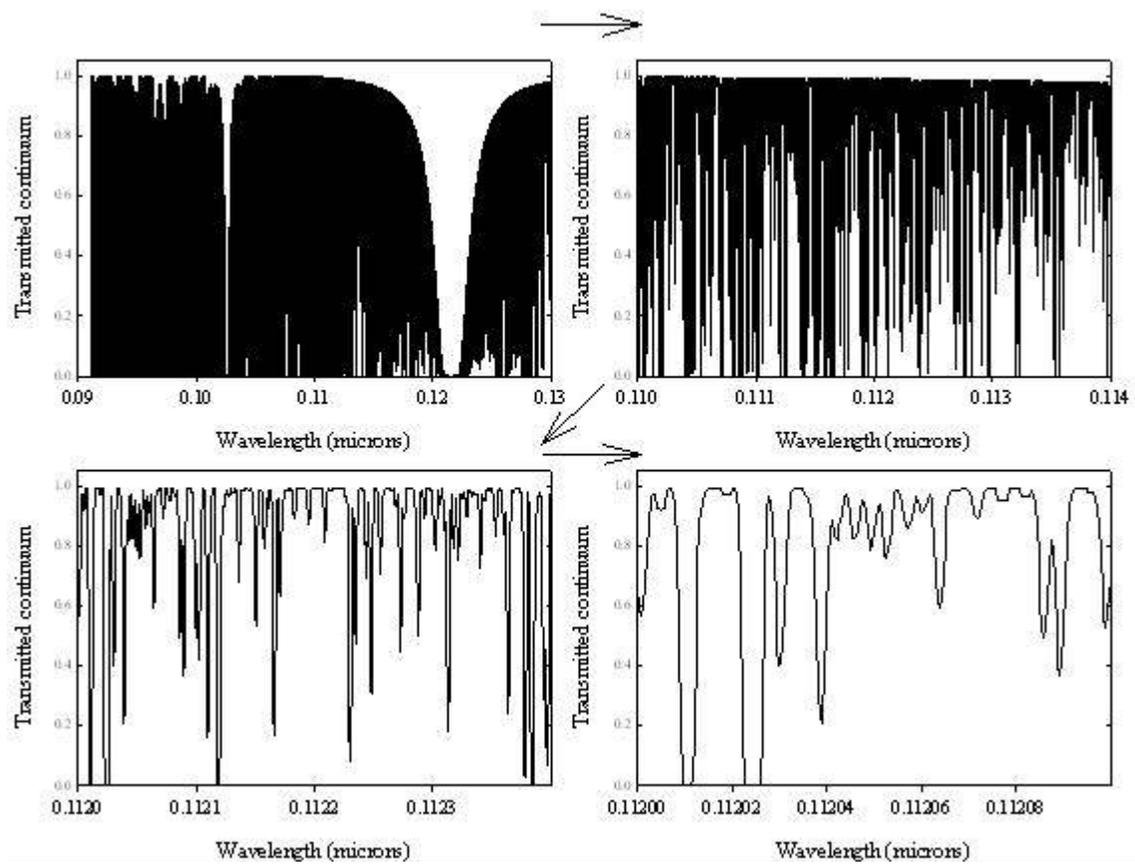

Figure 20(b). Fine line and continuum overlap for the TH85 PDR at the same point as Figure 20(a).

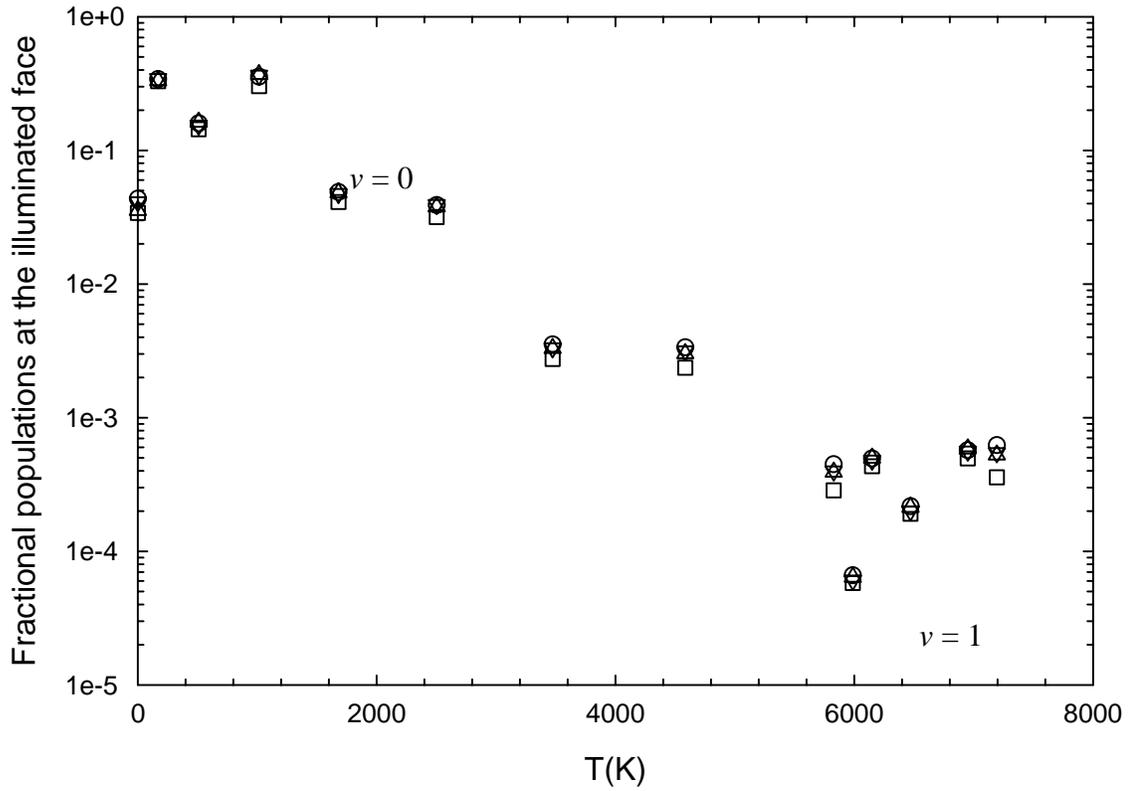

Figure 21. Fractional populations at the illuminated face of model F1 of Leiden PDR workshop (2004). The open circles represent our results. The triangle down, triangle up and squares represent results of Le Bourlot et al. (1995), Sternberg (1999), and Black et al. (1987) respectively.

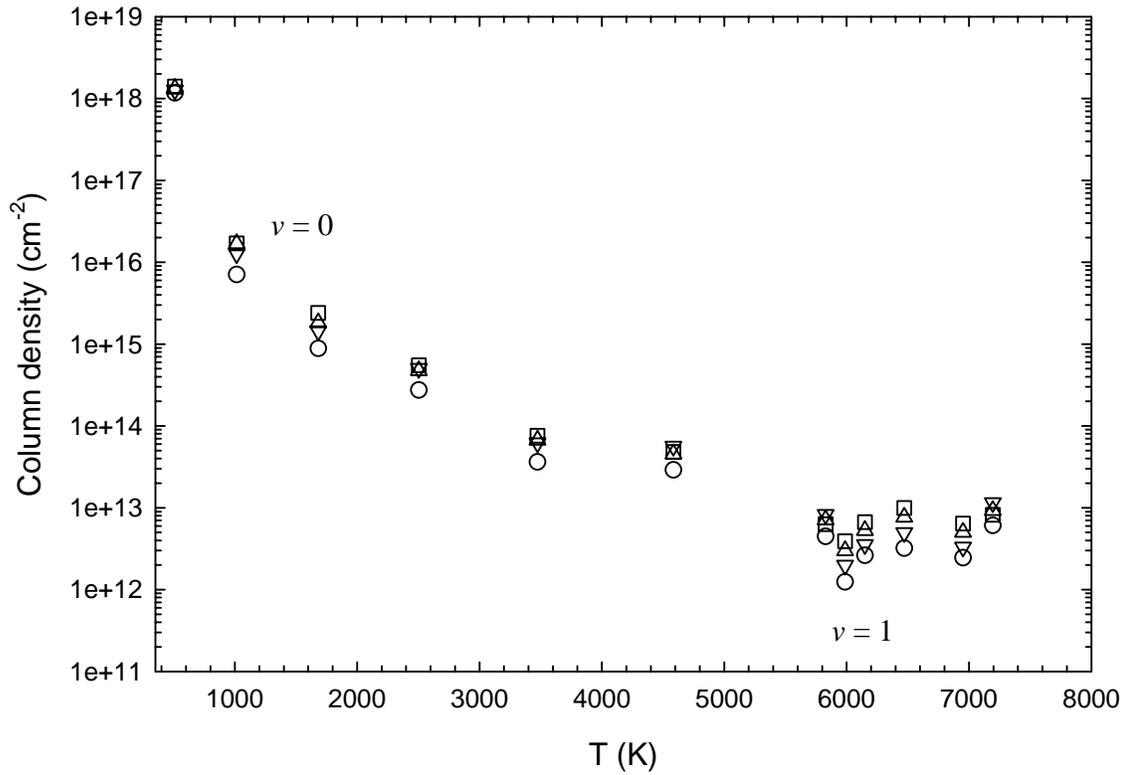

Figure 22. Integrated column density of model F1 of Leiden PDR workshop (2004). The open circles represent our results. The triangle down, triangle up and squares represent results of Le Bourlot et al. (1995), Sternberg (1999), and Black et al. (1987) respectively.

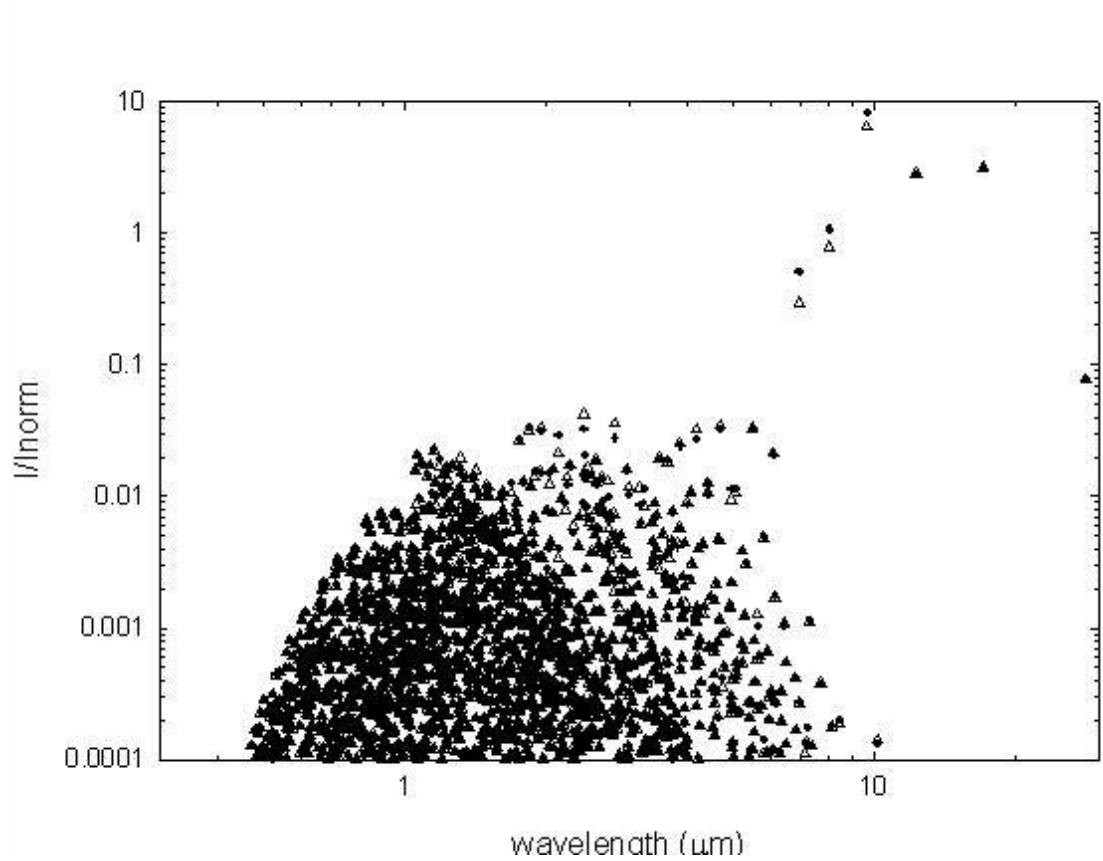

Figure 23. A Comparison of resulting spectra at 500K with (black) and without (white) Gaussian random noise in collisional rates.

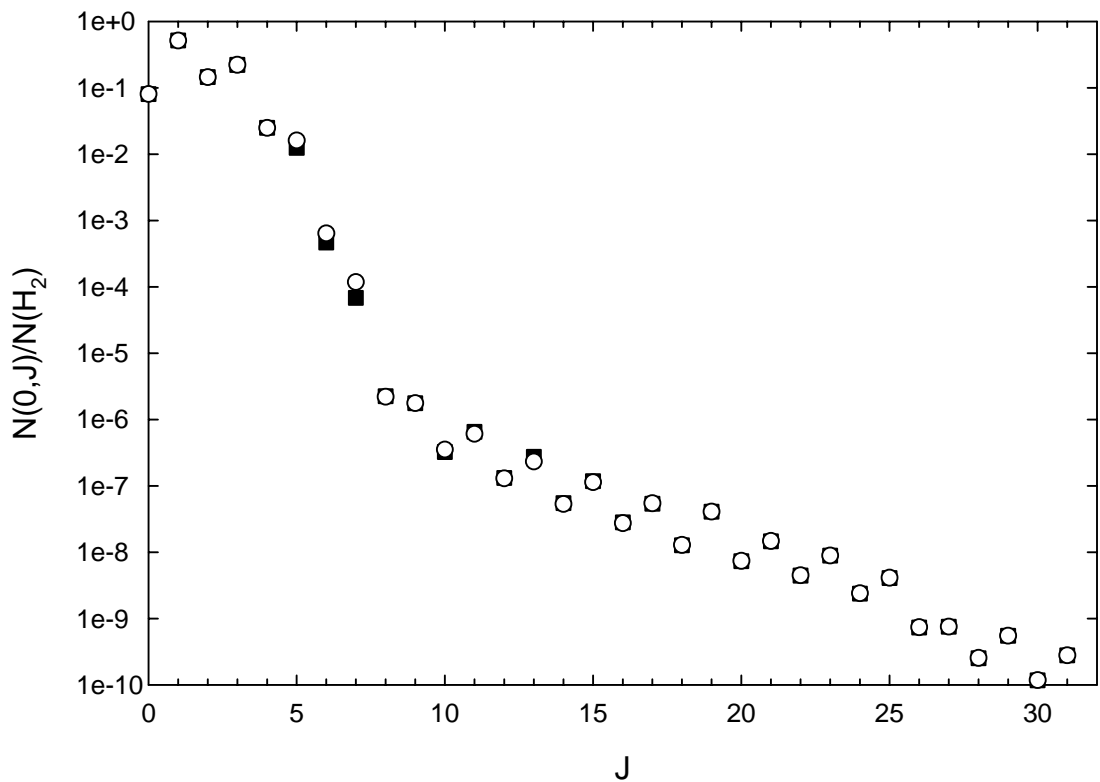

Figure 24. The level populations for *v*=0 and *T*=500 K with and without the random Gaussian noise in collisional data. The rotation quantum number *J* is indicated along the x-axis while the y-axis gives the populations per $H_2$. The black squares and open circles represent populations with and without noise.